\documentclass[aps,prd,twocolumn,showpacs,superscriptaddress,groupedaddress,nofootinbib]{revtex4-1}  
\usepackage[dvips]{graphicx}
\usepackage{lineno}

\usepackage{amsmath}
\usepackage{amssymb}
\usepackage{amsfonts}
\usepackage{amsthm}

\usepackage{xcolor}
\definecolor{red}      {rgb}{0.8,0.0,0.0}
\definecolor{green}    {rgb}{0.0,0.6,0.0}
\definecolor{darkblue} {rgb}{0.0,0.1,0.7}
\definecolor{brown}    {rgb}{0.6,0.1,0.0}
\definecolor{gray}     {rgb}{0.6,0.6,0.6}
\definecolor{darkgreen}{rgb}{0.0, 0.545098, 0.0}
\definecolor{orange}   {RGB}{238,80,25}
\definecolor{purple}   {rgb}{0.5,0.0,0.5}
\definecolor{babypink} {rgb}{0.64, 0.44, 0.44}

\begin{document}

\title{Equation of State in 2+1 Flavor QCD at High Temperatures}

\author{A. Bazavov$^a$, P. Petreczky$^b$, J. H. Weber$^{c,d}$}
\affiliation{
$^a$ Department of Computational Mathematics, Science and Engineering
and Department of Physics and Astronomy,
Michigan State University, East Lansing, MI 48824, USA
$^b$ Physics Department, Brookhaven National Laboratory, Upton, New York 11973, USA\\
$^c$ Physik Department, Technische Universit\"{a}t M\"{u}nchen, D-85748 Garching, Germany\\
$^d$ Exzellenzcluster Universe, Technische Universit\"{a}t M\"{u}nchen, 
D-85748 Garching, Germany
}

\begin{abstract}
We calculate the Equation of State at high temperatures in 2+1 flavor QCD 
using the highly improved staggered quark (HISQ) action. 
We study the lattice spacing dependence of the pressure at high temperatures
using lattices with temporal extent $N_{\tau}=6,~8,~10$ and $12$ and perform
continuum extrapolations. 
We also give a continuum estimate for the Equation of State up to temperatures 
$T=2$\,GeV, which are then compared with results of the weak-coupling 
calculations. 
We find a reasonably good agreement with the weak-coupling calculations at 
the highest temperatures.

\end{abstract}

\maketitle

\section{Introduction}\label{intro}

Over the last several years there was a focused effort to calculate the Equation 
of State of strongly interacting matter at net zero baryon density 
in lattice QCD using physical or nearly 
physical quark masses and improved staggered action \cite{Aoki:2005vt, Bernard:2006nj,Cheng:2007jq,Bazavov:2009zn,Cheng:2009zi,Borsanyi:2010cj, Borsanyi:2013bia,Bazavov:2014pvz}. 
As the result the continuum extrapolated Equation of State (EoS) has been 
obtained in 2+1 flavor QCD for physical light and strange quark masses 
\cite{Borsanyi:2013bia,Bazavov:2014pvz}. 
The calculations have been performed using two different improved staggered 
discretization schemes, the so-called stout action and the highly improved 
staggered quark (HISQ) action. 
These calculations cover a temperature range up to $T=400-500$\,MeV. 
Overall the results of these calculations agree well, except for the highest 
temperatures, where tension between the two results can be seen 
\cite{Bazavov:2014pvz}.
It is important to clarify if this tension is just due to some statistical
fluctuations or part of a systematic trend. 
Furthermore, for the comparison with the weak-coupling results it is highly desirable to extend the EoS calculations to higher temperatures. 
At temperatures $T>400$ MeV the charm quark contributes significantly to 
thermodynamic quantities and has to be included in the calculations 
\cite{Borsanyi:2016ksw}. 
Thus, one has to perform the calculations of the EoS in 2+1+1 flavor QCD. 
However, in the weak-coupling calculations the inclusion of the charm quark 
complicates the analysis, and the effects of the charm 
quark on the thermodynamic quantities are only known up to next-to-leading 
order (NLO) \cite{Laine:2006cp}. Also it is more difficult to control the discretization effects 
in the presence of the charm quark due to its large mass.
Therefore, for comparison of the lattice QCD results and the weak-coupling 
results it is advantageous to consider thermodynamic quantities in 2+1 flavor 
QCD at higher temperatures. 
Such calculations also provide a solid reference point for estimating the 
charm quark contribution to QCD thermodynamics. 

The purpose of this work is to extend the calculations presented in Ref. \cite{Bazavov:2014pvz} to higher temperatures. 
As in Ref. \cite{Bazavov:2014pvz} the HISQ action will be used together 
with the physical value of the strange quark mass.
The lattice spacing (cutoff) dependence of the pressure will be studied in 
detail.
In the previous studies the continuum extrapolations have been performed for 
the trace anomaly; 
the pressure and other thermodynamic properties have been obtained from the 
trace anomaly using the integral method \cite{Boyd:1996bx}.
The cutoff dependence of the trace anomaly, however, is expected to be more
complicated than the cutoff dependence of the pressure. The reason for this
is the following. In the weak-coupling picture the trace anomaly receives 
contributions starting at three loop, i.e. at order $\alpha_s^2$. Therefore,
the understanding of the cutoff dependence of the trace anomaly at high temperature
would in principle require a three-loop calculation in lattice perturbation theory.
This is clearly formidable task. On the other hand the pressure at high temperature
receives the leading contribution at one loop (${\cal O}(\alpha_s^0)$)
corresponding to the ideal gas limit.
Therefore, the cutoff dependence of the pressure at high temperature is known and to fairly 
good approximation is described by the free gas \cite{Beinlich:1995ik,Heller:1999xz}.

For better understanding of the cutoff dependence of the EoS at high 
temperatures and a better control of the continuum extrapolation it is 
desirable to study the cutoff dependence of the pressure directly. 
This may also help to understand the difference between the continuum-extrapolated results and the results obtained with p4 or asqtad-improved 
staggered actions and $N_{\tau}=6$ and $8$ at high temperatures 
\cite{Cheng:2007jq,Bazavov:2009zn} since cutoff effects here should be small.

It is expected that thermodynamic properties are not sensitive to the value of the 
light quark masses at high temperatures. The quark mass dependence of the EoS was studied in
Ref. \cite{Borsanyi:2010cj} and it was found that for light quark masses smaller than $0.4m_s$
the quark mass dependence is very small for $T>450$ MeV. Therefore,  we consider light quark masses which 
are five times smaller than the strange quark mass, $m_l=m_s/5$, instead of 
the physical value.
This choice of the light quark mass corresponds to a pion mass of about $320$\,MeV in the continuum limit.

The rest of the paper is organized as follows. 
In Section II we discuss details of the lattice calculations.
In Section III we show our results for the trace anomaly. 
In Section IV we present the calculation of the pressure and its cutoff 
dependence. 
Comparison of the lattice calculations to the weak-coupling results is discussed
in Section V. 
Finally Section VI contains our conclusions. 
Some technical aspects of the calculations are presented in the appendices.

\section{Lattice calculations at zero temperature}

The goal of this paper is to extend the calculations of the QCD Equation 
of State in Ref. \cite{Bazavov:2014pvz} to higher temperatures.
Therefore, as in Ref. \cite{Bazavov:2014pvz} we use tree-level improved gauge action
and HISQ action for quarks.
To calculate the EoS gauge configurations at zero temperature had to be 
generated to perform the subtraction of the UV divergences in the 
thermodynamic quantities as well for the determination of the lattice spacing. 
We generated the gauge configurations at $T=0$ using the rational hybrid 
Monte-Carlo (RHMC) algorithm at five values of the lattice gauge coupling
$\beta=10/g^2$. 
The parameters of the simulations are shown in Tab. \ref{tab:T0}, including 
the lattice volume.
The lowest two $\beta$ values will be used for the purpose of comparison with 
the previous 2+1 flavor results at smaller light quark masses \cite{Bazavov:2014pvz}, enabling us to 
quantify the quark mass effects in the scale setting procedure as well as in 
the thermodynamic
quantities. 
\begin{table}
\begin{tabular}{ccccc}
\hline
$\beta$ &  $m_s$  &  vol    &  a [fm]  &  \# traj.  \\
\hline
7.030   & 0.03560 & $48^4$  &  0.08253 &  1890 \\
7.825   & 0.01542 & $64^4$  &  0.04036 &  1265 \\
8.000   & 0.01299 & $64^4$  &  0.03469 &  3927 \\
8.200   & 0.01071 & $64^4$  &  0.02924 &  3927 \\
8.400   & 0.00887 & $64^4$  &  0.02467 &  3927 \\
\hline
\end{tabular}
\caption{The parameters of the $T=0$ simulations.}
\label{tab:T0}
\end{table}

The lattice spacings corresponding to the highest three $\beta$ values in 
Tab. \ref{tab:T0} are smaller than $0.035$ fm. 
At these small lattice spacings it is expected that the Monte-Carlo (MC) 
evolution of the topological charge will effectively freeze.
Indeed, we observe that the topological charge does not change in the MC 
evolution. 
To deal with this problem we generated MC streams corresponding to different 
values of topological charge, namely $Q=0,~1$ and $2$. 
We checked whether the observables of interest are sensitive to the value of 
the topological charge, but we did not find any sensitivity. 
The dependence of different observables on the topological charge is discussed 
in Appendix \ref{appA}. 

To determine the lattice spacing we calculated the static quark anti-quark 
potential. 
The lattice spacing is determined through the scale parameters $r_1$ and 
$r_2$ defined as

\begin{equation}
r^2 \left . \frac{d V(r)}{d r} \right|_{r=r_1}=1,~~ 
r^2 \left .\frac{d V}{d r}\right |_{r=r_2}=\frac{1}{2}.
\end{equation}
The parameter $r_1$ is widely used by the MILC and HotQCD collaborations to set 
the lattice spacing (see e.g. Ref. \cite{Bazavov:2014pvz}). 
The value of this parameter is $r_1=0.3106$ fm \cite{Bazavov:2010hj}. 
Since we consider smaller lattice spacings it is useful to consider the scale 
parameter $r_2$. 
The calculation of the static potential and the determination of $r_1$ and 
$r_2$ scales is discussed in Appendix \ref{appA}.

For the two lower $\beta$ values in Tab. \ref{tab:T0} we could compare the 
results on the static potential calculated for $m_l=m_s/5$ with the previous 
calculations performed at $m_l=m_s/20$ to study quark mass effects.
We find no quark mass effects at the shortest distances. 
Quark mass effects increase with increasing distances but are less then 
$0.2\%$ for $r<r_1$. 
At distances around $r=r_1$ the statistical errors in the static potential are 
large enough so that no quark mass effects in the derivative of the potential 
can be seen. 
Therefore, we can combine the newly determined values of $r_1$ with the 
previously published HotQCD results to obtain $r_1/a$ as function of $\beta$.
The details of this analysis are given in Appendix \ref{appA}.

\section{The QCD trace anomaly}\label{secTmumu}

\begin{figure*}
\includegraphics[width=8cm]{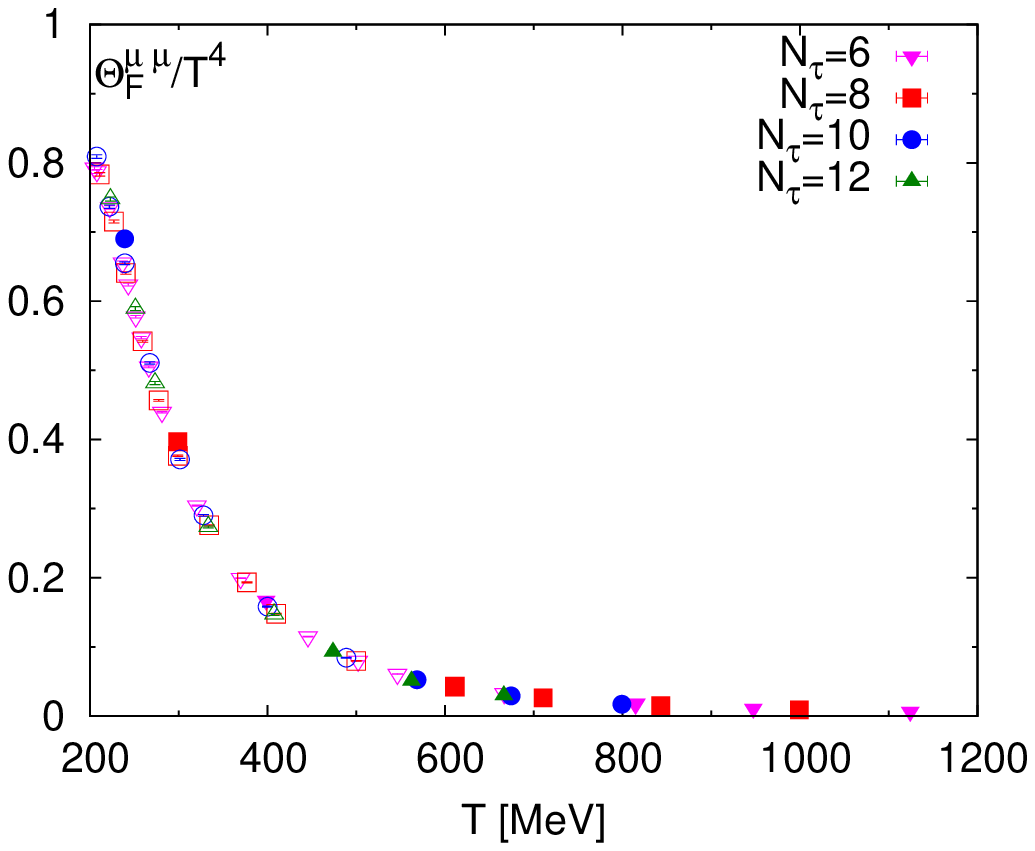}
\includegraphics[width=8cm]{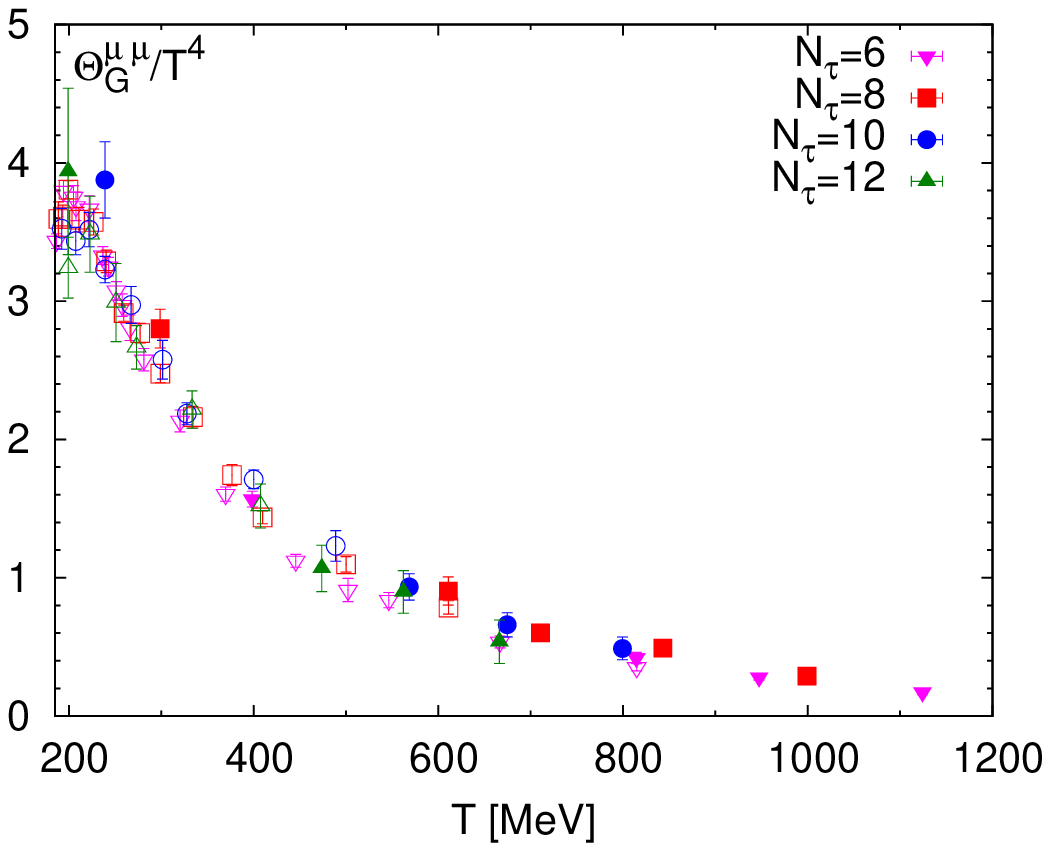}
\caption{The fermionic part (left) and the gauge part (right) of the trace anomaly obtained with HISQ action.
The open symbols correspond to the HotQCD results for $m_l=m_s/20$ \cite{Bazavov:2014pvz}.}
\label{fig:e3p_GF}
\end{figure*}

To extend the calculation of the Equation of State of 2+1 flavor QCD we used 
the integral method, which relies on the calculation of the trace of the 
energy momentum tensor $\Theta^{\mu \mu}=\epsilon-3p$ or the trace anomaly 
for short \cite{Cheng:2007jq,Bazavov:2009zn}.
The pressure can be calculated in terms of the trace anomaly as follows:

\begin{equation}
\frac{p(T)}{T^4}-\frac{p(T_0)}{T_0^4}=\int_{T_0}^T dT' 
\frac{\epsilon-3p}{T'^5},
\label{int}
\end{equation}
where $T_0$ is some reference temperature, which is sufficiently small, so 
$p(T_0)$ can be either set to zero or taken from the hadron resonance gas 
calculation \cite{Cheng:2007jq,Bazavov:2009zn,Bazavov:2014pvz}.
The trace anomaly can be expressed in terms of the expectation values of the 
gauge action, $\langle s_G\rangle_{\tau (0)}$, and the light, 
$\langle \bar{\psi}\psi \rangle_{l,\tau (0)}$, and strange,
$\langle \bar{\psi}\psi \rangle_{s,\tau (0)}$, quark condensates,  
calculated at finite and 
zero temperature, respectively. 
For the HISQ action the corresponding formula has the 
form \cite{Bazavov:2014pvz}:

\begin{align}
\displaystyle
\frac{\epsilon-3p}{T^4} 
&\equiv
\displaystyle
\frac{\Theta^{\mu\mu}_G(T)}{T^4} +
\frac{\Theta^{\mu\mu}_F(T)}{T^4} \; , \\[2mm]
\displaystyle
\frac{\Theta^{\mu\mu}_G(T)}{T^4}
&=
R_\beta
\left[ \langle s_G \rangle_0 - \langle s_G \rangle_\tau \right] N_\tau^4 \;, 
\label{e3pG}  \\[2mm]
\displaystyle
\frac{\Theta^{\mu\mu}_F(T)}{T^4}  
&= - R_\beta R_{m} [
2 m_l\left( \langle\bar{\psi}\psi \rangle_{l,0}
- \langle\bar{\psi}\psi \rangle_{l,\tau}\right)  \nonumber \\[2mm] 
\displaystyle
&\phantom{=-}+ m_s \left(\langle\bar{\psi}\psi \rangle_{s,0}
- \langle\bar{\psi}\psi \rangle_{s,\tau} \right )
 ] N_\tau^4 \; .
\label{e3pF}
\end{align}

Here we used the same notation as in Ref. \cite{Bazavov:2014pvz} and we made 
explicit the separation of the trace anomaly into the fermionic and gluonic 
parts. 
Furthermore, we introduced the nonperturbative beta function and mass 
renormalization function
defined as \cite{Cheng:2007jq,Bazavov:2009zn}

\begin{align}
\displaystyle
R_{\beta}(\beta) 
&= 
\frac{r_1}{a} \left( {{\rm d} (r_1/a) \over {\rm d} \beta} \right)^{-1}\; ,\label{Rbeta}\\
\displaystyle
R_m(\beta) 
&= 
\frac{1}{m_s(\beta)}
\frac{{\rm d} m_s(\beta)}{{\rm d}\beta}\; .
\label{Rm}
\end{align}

The calculation of the nonperturbative beta function is discussed in 
Appendix \ref{appA}. 
The mass renormalization function is taken from Ref. \cite{Bazavov:2014pvz}. 
As also discussed in Appendix \ref{appA}, the new zero temperature 
calculations are consistent with this mass renormalization function.

To calculate the trace anomaly at temperatures corresponding to the values 
of $\beta$ given in Table \ref{tab:T0} we use the finite temperature gauge configurations from the TUMQCD collaboration \cite{Bazavov:2016uvm, 
TUM-EFT81/16}.
These gauge configurations have been generated on 
$N_{\sigma}^3 \times N_{\tau}$ lattices with $N_{\tau}=4,~6,~8,~10$ and $12$ 
and $N_{\sigma}=4 N_{\tau}$. 
The maximal temperature corresponding to these lattices is about $2$\,GeV. 

\begin{figure}
\includegraphics[width=8cm]{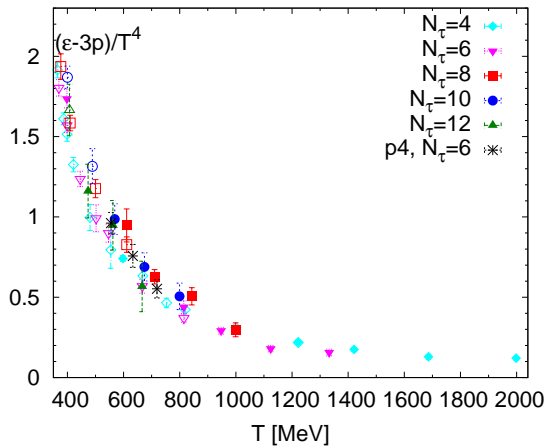}
\caption{The numerical results for the trace anomaly 
at two different quark masses (see text).
The open symbols correspond to $m_l/m_s=1/20$, while the filled symbols
correspond to $m_l/m_s=1/5$.
The bursts correspond to calculations with p4 action and $m_l=m_s/10$
\cite{Cheng:2007jq}.}
\label{fig:e-3p_high}
\end{figure}

Now we will discuss our numerical results on the trace anomaly,
in particular, its dependence on the light quark masses. There are
two sources of quark mass dependence of the trace anomaly.
First, is the dependence of the trace anomaly on the light sea quark masses.
The second is the explicit dependence of the fermionic part of
the trace anomaly on the light quark mass. As we will see later
there are also differences in the cutoff ($N_{\tau}$) dependence of
the fermionic and gluonic parts of the trace anomaly. Therefore,
in the following we will discuss the numerical results for $\Theta_F^{\mu \mu}$
and $\Theta_G^{\mu \mu}$ separately.
The fermionic part of the trace anomaly, $\Theta_F^{\mu \mu}$ 
is shown in Fig. \ref{fig:e3p_GF} (left)
and compared with the published HotQCD results obtained for $m_l=m_s/20$ \cite{Bazavov:2014pvz}
and shown as open symbols.
To take into account the explicit dependence on the light
quark masses in the calculation of 
$\Theta_F^{\mu \mu}$ we used the value $m_l=m_s/20$ instead of $m_l=m_s/5$. 
We see from the figure that after adjusting the light quark mass there is no
quark mass dependence in $\Theta^{\mu \mu}_F$ for $T>300$\,MeV, i.e. the
the quark mass dependence of $\Theta_F^{\mu \mu}$ due to the sea quarks is very small.
From Fig. \ref{fig:e3p_GF} (left) we also see that 
the cutoff effects in $\Theta_F^{\mu \mu}$ are very small in accordance with the
previous study \cite{Bazavov:2014pvz}. Finally, we note
that statistical errors for $\Theta_F^{\mu \mu}$ are tiny.
Our results for the gluonic part of the trace anomaly, $\Theta_G^{\mu \mu}$, are shown in Fig. \ref{fig:e3p_GF}(right).
The cutoff and quark mass dependence of $\Theta_G^{\mu \mu}$ can be clearly seen.
The quark mass dependence of $\Theta_G^{\mu \mu}$ is due to the sea quarks and thus cannot
be corrected. It is the sole source of the quark mass dependence of the trace anomaly shown in Fig. \ref{fig:e-3p_high}.
We see, however, that quark mass effects become smaller at high temperatures
and statistically are not significant for $T>400$ MeV. The statistical errors for $\Theta_G^{\mu \mu}$ 
are much larger than for $\Theta_F^{\mu \mu}$ and it is the dominant contribution to the trace
anomaly. 

For the calculation of the trace anomaly at high temperatures we also used
$N_{\tau}=4$ lattices from the TUMQCD collaboration obtained with $m_l=m_s/20$ 
\cite{Bazavov:2016uvm} and $m_l=m_s/5$ \cite{TUM-EFT81/16}.
Our results for the trace anomaly at high temperatures are summarized in
Fig. \ref{fig:e-3p_high}.
The open symbols in the figure refer to
$m_l/m_s=1/20$ results, while the filled symbols refer to $m_l/m_s=1/5$ results.
All the $m_l=m_s/20$ results for $\epsilon-3p$ are from Ref. \cite{Bazavov:2014pvz},
except the ones for $N_{\tau}=4$ and those for $N_\tau=6$ with $\beta=7.03$ or $7.825$. 
>From Fig. \ref{fig:e-3p_high}
we see that $m_s/20$ results smoothly match to the $m_s/5$ results at high temperatures.
This is expected. From the calculations of the trace anomaly performed with stout action
at several quark masses we can estimate that the difference in the trace anomaly calculated
for $m_l=m_s/5$ and $m_l=m_s/20$ is $10\%,~4\%,~3\%$ and $<1\%$ for $T=300,~400,~500$ and $600$ MeV,
respectively. Our calculations with $m_l=m_s/5$ at $T \le 400$ MeV confirm these expectations.
The statistical errors shown in Fig. \ref{fig:e-3p_high} are much larger than the above
differences in the temperature range of interest, 
so no quark mass effects are visible given the errors. In Fig. \ref{fig:e-3p_high}
we also show the trace anomaly calculated with
p4 action  
for $m_l/m_s=1/10$ and $N_{\tau}=6$ \cite{Cheng:2007jq}. The corresponding results agree well
with the HISQ results. Overall we see that quark mass effects are very small at high temperatures
and therefore it is justified to study QCD thermodynamics with $m_l=m_s/5$ in this region.
Finally, we note that there is no visible cutoff dependence for $\epsilon-3p$ for $N_{\tau}\ge 8$
in the high temperature region, while the $N_{\tau}=4$ and $6$ data are systematically below
the $N_{\tau}\ge 8$ results.

\begin{figure}
\includegraphics[width=8cm]{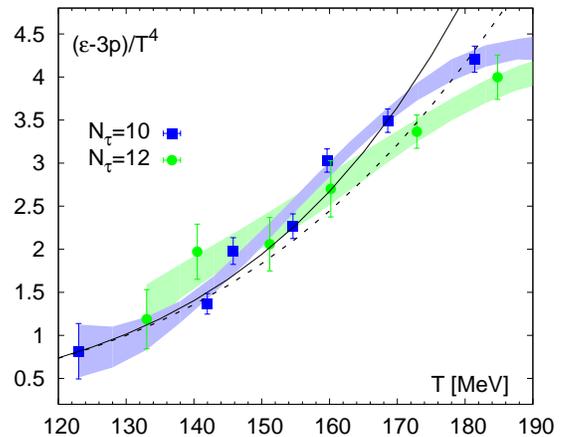}
\caption{The trace anomaly calculated in the low temperature 
region using $N_{\tau}=10$ and $N_{\tau}=12$ lattices. 
The bands correspond to interpolations (see text). 
The dashed line corresponds to HRG-PDG, while the solid line corresponds to 
HRG-QM.}
\label{fig:e-3p_low}
\end{figure}

While the main purpose of this work is to extend the EoS calculations to high 
temperatures we also revisited the trace anomaly in the low temperature 
region for $N_{\tau}=10,~12$ using the gauge configurations generated by the 
TUMQCD collaboration for the study of the Polyakov loop 
\cite{Bazavov:2016uvm}. The reason behind this is the fact that unlike
in Ref. \cite{Bazavov:2014pvz} the continuum extrapolations  will
be performed in terms of the pressure and not the trace anomaly.
Therefore, a more accurate determination of the pressure and the trace anomaly
at low temperatures is needed.
We added the following temperatures: $T=123$\,MeV ($N_{\tau}=10$), and 
$T=133$\,MeV and $T=140$\,MeV ($N_{\tau}=12$). 
The $T=0$ gauge configurations are the same as in Ref. \cite{Bazavov:2014pvz}. 
The numerical results for the trace anomaly in the low temperature region are 
shown in Fig. \ref{fig:e-3p_low}.
We performed interpolations of the lattice results on $\epsilon-3p$ using 
smoothing splines. 
The number of knots in the spline and the value of the smoothing parameter 
have been adjusted such that we obtain a smooth behavior with minimum 
number of knots and keep the $\chi^2/{\rm df}$ close to one. 
The statistical error on the spline has been estimated using bootstrap method. 
We see sizable differences in $\epsilon-3p$ calculated with $N_{\tau}=10$ and 
$N_{\tau}=12$, indicating residual cutoff effects in the region 
$160$\,MeV $<T< 180$\,MeV. 
We also compare our results with the hadron resonance gas (HRG) model. 
We show two versions of the HRG model: one that takes into account all states 
from the particle data group, which we label as HRG-PDG, and one that includes 
baryon states that are not yet discovered experimentally, but predicted by the quark model (missing states). 
We label the latter model as HRG-QM. 
The details of the HRG models are described in Appendix \ref{appB}. 
There we also introduce the HRG-QM models for non-zero lattice spacing in
addition to the continuum HRG-QM model shown in \ref{fig:e-3p_low}.
>From the figure we see that the difference between the two HRG models is 
only significant for $T>150$\,MeV. 
The lattice results for $N_{\tau}=10$ and $12$ agree with the HRG models only 
for $T<145$\,MeV. 
This is in agreement with the previous results \cite{Borsanyi:2013bia, 
Bazavov:2014pvz}.
Unlike in Ref. \cite{Bazavov:2014pvz} we did not require that the 
interpolations agree with the HRG model at low temperatures. 
So these results serve as independent check for the validity of the HRG model.

\section{The pressure of 2+1 flavor QCD from low to high temperatures}\label{secp}

In this section we discuss the calculation of the pressure in the wide 
temperature range from $T=120$\,MeV to $2000$\,MeV. 
For this purpose we combine the published HotQCD results for the trace 
anomaly with the results obtained for $m_l=m_s/5$ and discussed in the 
previous section. 
We use the published HotQCD results for $T \le 407$\,MeV and $N_{\tau}=12$, $T \le 489$\,MeV 
and $N_{\tau}=10$, $T \le 611$\,MeV and $N_{\tau}=8$, and $T \le 815$\,MeV and 
$N_{\tau}=6$ \cite{Bazavov:2014pvz}. 
For temperatures higher than these we use the new $m_s/5$ results with 
$N_{\tau}=12,~10,~8$ and $6$. 
Since the quark mass effects are smaller than the statistical errors we treat 
these two data sets as one and perform interpolations of the data from
the combined set. 
Using the resulting interpolating function we can calculate the pressure 
according to Eq. \eqref{int}. 
Essentially we will be computing the pressure for lines of constant physics
corresponding to $m_l=m_s/20$ even though the data for the trace anomaly at the high
temperatures come from calculations at $m_l=m_s/5$. 
To fix the pressure completely we need to specify the lower integration 
limit $T_0$ as well as the value of the pressure at $T=T_0$. 
The lower integration limit $T_0$ is determined by the lowest data point 
for which a lattice calculation of $\epsilon-3p$ is available for given 
$N_{\tau}$. As in Ref. \cite{Bazavov:2014pvz} we will use HRG to estimate
the pressure at $T_0$. However,
when choosing the value of $p(T_0)$ we need to take into account the 
discretization effects of the staggered fermion formulation due to the 
distortion of the hadron spectrum. 
Therefore we calculate $p(T_0)$ in the HRG model with distorted hadron 
spectrum. 
The details of these calculations are discussed in Appendix \ref{appB}. 
As the result we obtain a value $p(T_0,N_{\tau})$ for each $N_{\tau}$. 
The values of $T_0$ and $p(T_0,N_{\tau})$ used in the calculation of the 
pressure are given in Table \ref{tab:p0}.
\begin{figure*}
\includegraphics[width=8cm]{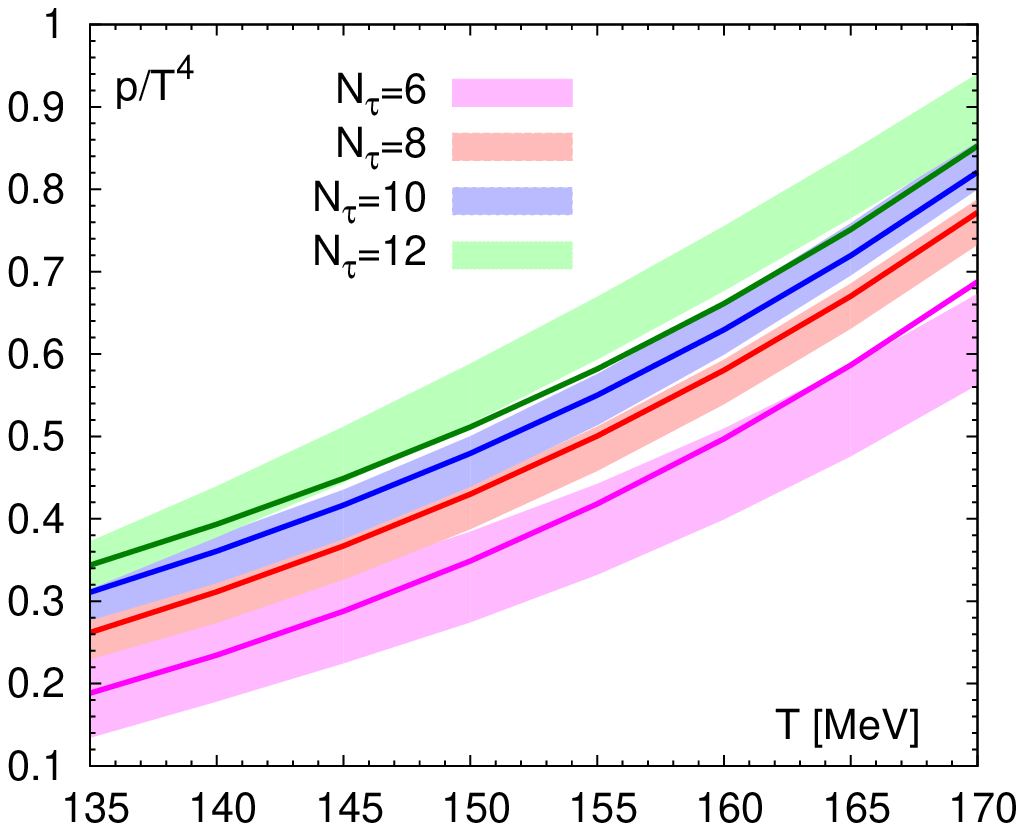}
\includegraphics[width=8cm]{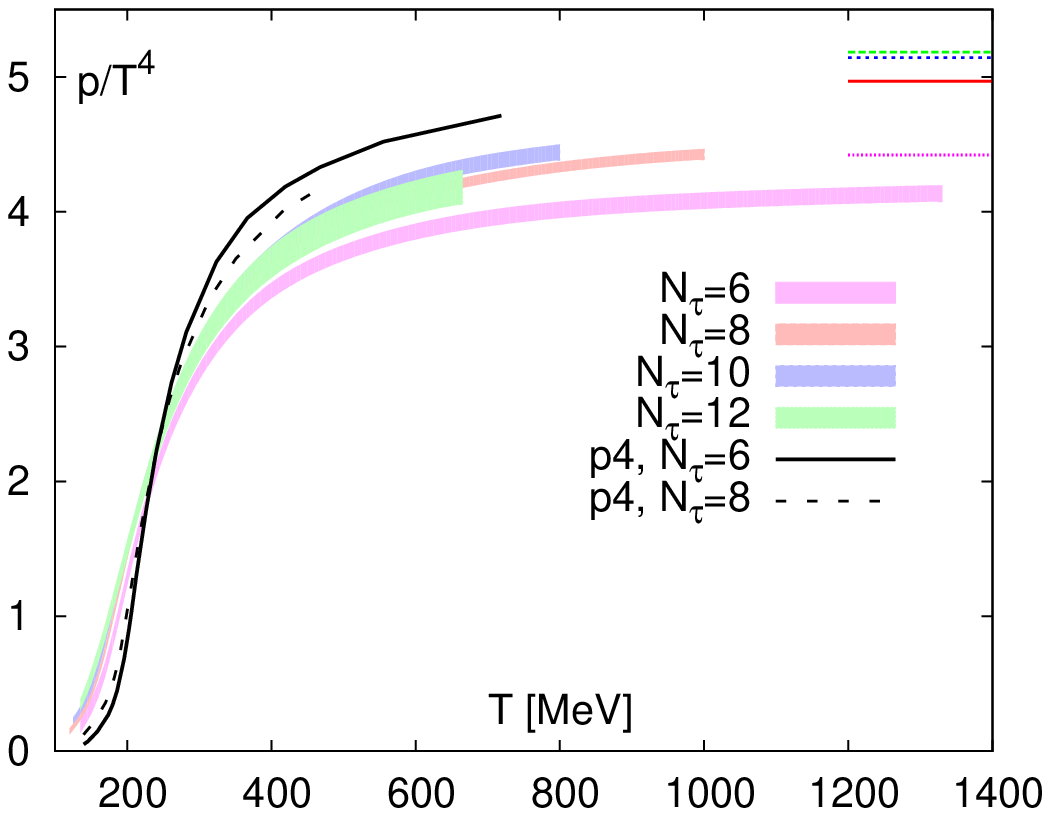}
\caption{Left:The pressure in the low temperature region.
The lines correspond to HRG with distorted hadron spectrum (see text).
Right: the pressure in the entire temperature range.
The horizontal lines correspond to the free theory result.
Also shown are the results for the pressure obtained with p4 action and
$N_{\tau}=6$ or $8$ \cite{Cheng:2007jq,Bazavov:2009zn}.
}
\label{fig:p_all}
\end{figure*}

\begin{table}
\begin{tabular}{|ccc|}
\hline
$N_{\tau}$ & $T_0$ [MeV] & $p(T_0,N_{\tau})$ \\
\hline
6          & 135         & 0.189(54)         \\
8          & 120         & 0.145(22)         \\
10         & 125         & 0.226(23)         \\
12         & 135         & 0.344(29)         \\
\hline
\end{tabular}
\caption{The values of $T_0$ and $p(T_0)$ used to calculate
the pressure for different $N_{\tau}$ (see text).}
\label{tab:p0}
\end{table}
With these inputs we can calculate the pressure for $N_{\tau}=6,~8,~10$ and $12$.
The results are shown in Fig. \ref{fig:p_all}. 
We see significant cutoff dependence in the low temperature region and 
smaller cutoff dependence in the high temperature region. 
In the low temperature region the pressure follows qualitatively the cutoff 
dependence obtained in the HRG model with distorted hadron spectrum, 
cf. Fig. \ref{fig:p_all} (left). 
The continuum limit for the pressure is approached from below. 
The pressure shows stronger cutoff dependence than the trace anomaly. 
Both of these features could be understood in the framework of the 
HRG model with distorted hadron spectrum (see Appendix \ref{appB}).

\begin{figure}
\includegraphics[width=8cm]{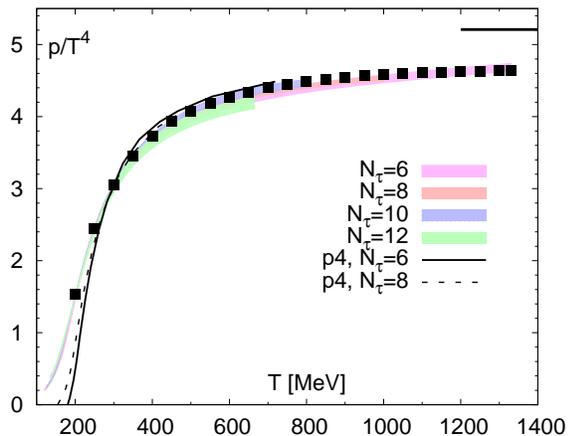}
\caption{
The pressure calculated with HISQ action for different $N_{\tau}$ and 
corrected for cutoff effects. 
The filled squares are the continuum results for the pressure.
For comparison we also plot the p4 results for the pressure corrected for 
cutoff effects at high temperatures.
}
\label{fig:p_cont}
\end{figure}

At high temperatures the cutoff dependence of the pressure can be understood
in the weak-coupling picture. In this picture the pressure can be written
as the sum of quark and gluon pressures with the latter being defined as the QCD 
pressure for $N_f=0$\footnote{Note that this decomposition of the pressure
into the quark and gluon pressures is different from the decomposition of
$\Theta^{\mu \mu}$ into $\Theta_G^{\mu \mu}$ and $\Theta_F^{\mu \mu}$.
The quark pressure does not vanish for zero quark mass but $\Theta_F^{\mu \mu}$
does.}. The cutoff dependence of the quark and gluon pressures has been studied
in lattice perturbation theory up to order $\alpha_s$ \cite{Beinlich:1997ia,Heller:1999xz,Hegde:2008nx}. 
To a good approximation
this cutoff dependence is described by the ideal gas result. The cutoff dependence
of the gluon pressure is very small ($<1\%$) for $N_{\tau}\ge 6$ if improved gauge
action is used \cite{Beinlich:1995ik}. This is confirmed by direct lattice numerical
study \cite{Beinlich:1997ia}. Therefore we neglect it here. The cutoff dependence
of the quark pressure was studied in Refs. \cite{Heller:1999xz,Hegde:2008nx} for improved
staggered actions, namely the
Naik action and p4 action. The cutoff dependence of
the quark pressure is much bigger than of the gluon pressure for $N_{\tau}\le 12$. At tree level the HISQ
action has the same cutoff dependence as the Naik action. Thus, the ideal gas limit
for the HISQ action is determined by the result for Ref. \cite{Heller:1999xz}.
The ideal gas limit for each $N_{\tau}$ is shown in Fig. \ref{fig:p_all}
as a horizontal line.
Our numerical results for the pressure at high temperatures shown
in Fig. \ref{fig:p_all} (right) follow the same trend in terms of cutoff
dependence as the free theory result.
The $N_{\tau}=12$ result appears to be an exception, though given the 
statistical errors the deviations from the trend is not very significant. 
At quantitative level the cutoff effects in the pressure are smaller than 
in the free field theory. 
This observation is in line with the cutoff dependence of the pressure in 
SU(3) gauge theory \cite{Boyd:1996bx} as well as cutoff effects of quark 
number susceptibilities (QNS) obtained with HISQ action 
\cite{Bazavov:2013uja,Ding:2015fca}.
For comparison we also show the pressure obtained with p4 action and 
$N_{\tau}=6$ or $8$ lattices \cite{Cheng:2007jq,Bazavov:2009zn}.
The corresponding results are significantly larger than the ones obtained 
with HISQ action at high temperatures and significantly lower at small 
temperatures. 
This is most likely due to the large cutoff effects related to 
taste-symmetry breaking for the p4 action (see discussions in Appendix B).

The cutoff dependence of the pressure obtained with HISQ action and p4
action closely resembles the cutoff dependence of quark number 
susceptibilities (QNS) defined as second and fourth derivatives of the 
pressure with respect to quark chemical potential, 

\begin{equation}
\chi_{2n}^q=\frac{\partial^{2n} p(T,\mu_q)}{\partial \mu_q^{2n}},~n=1,2,~q=l,s.
\end{equation}

Since the cutoff dependence of the quark contribution to the pressure and the 
cutoff dependence of QNS are similar, we could use the latter to correct for 
the cutoff dependence of the former. 
This can be done as follows. 
We write the pressure as the sum of the quark and gluon pressures\footnote{
This can be done if the weak-coupling picture holds at
high temperatures, as expected.} 
$p(T)=p^q(T)+p^g(T)$. 
Assuming that the gluonic pressure has negligible cutoff dependence (see
the above discussions) we can 
write

\begin{equation}
p(T)=p(T,N_{\tau})+corr(T,N_{\tau}),
\end{equation}

\noindent where $p(T,N_{\tau})$ is the pressure at fixed lattice spacing ($N_{\tau}$) and 

\begin{equation}
corr(T,N_{\tau})=p^q(T)\left( 1 - \frac{p^q(T,N_{\tau})}{p^q(T)} \right )
\end{equation}

\noindent is the correction factor due to discretization errors. Here $p^q(T)$ stands
for the quark pressure in the continuum limit, while $p^q(T,N_{\tau})$ is the 
quark pressure at non-zero lattice spacing, $a=(N_{\tau} T)^{-1}$. 
If we assume that the cutoff dependence of the quark pressure is the same as 
of the second order QNS, $\chi_2^l$, i.e.
\begin{equation}
\frac{p^q(T,N_{\tau})}{p^q(T)} \simeq \frac{\chi_2^l(T,N_{\tau})}{\chi_2^l(T)}
\end{equation}
we can use the results of 
Ref. \cite{Bazavov:2013uja} to obtain the correction provided we also have 
an estimate for continuum quark pressure $p^q(T)$. 
Lattice calculations show that the QCD pressure is below the ideal gas limit 
by about $15\%$ at high temperatures. 
Therefore, the ideal quark pressure provides a fair estimate for $p_q(T)$. 
Thus, we have an estimate for the correction.
We apply this correction to the pressure calculated for fixed $N_{\tau}$.
The results are shown in Fig. \ref{fig:p_cont}. 
We see from the figure that the pressure bands corresponding to different 
$N_{\tau}$ agree within errors, i.e. applying the corrections largely reduces
the $N_{\tau}$ dependence of the results. 
We also see that while the p4 results are still higher than the HISQ results
they agree within the statistical errors of the latter. 
The cutoff dependence of the pressure is understood because to a fairly good approximation
it is given by the cutoff dependence of the free quark gas. This is not the case
for the cutoff dependence trace anomaly, 
which would require a three-loop calculation as mentioned in Section \ref{intro}.
\begin{figure}
\includegraphics[width=8cm]{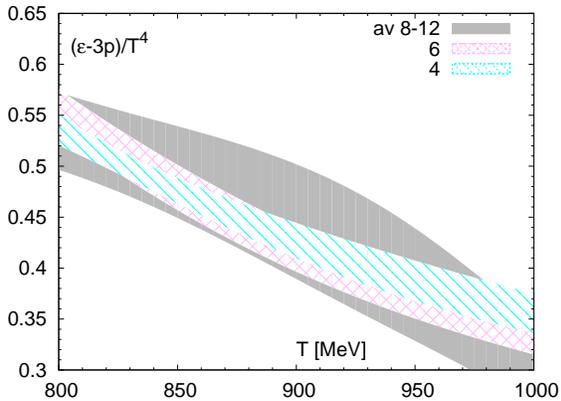}
\caption{
The interpolation of the lattice results for $(\epsilon-3p)/T^4$
for $N_{\tau}=8,~10$ and $12$ compared with the interpolations for
$(\epsilon-3p)/T^4$ obtained with $N_{\tau}=4$ and $6$.
The latter have been multiplied by $1.2$ and $1.4$ to bring them
into agreement with the former interpolation.}
\label{fig:e-3p_corr}
\end{figure}
\begin{figure}
\includegraphics[width=8cm]{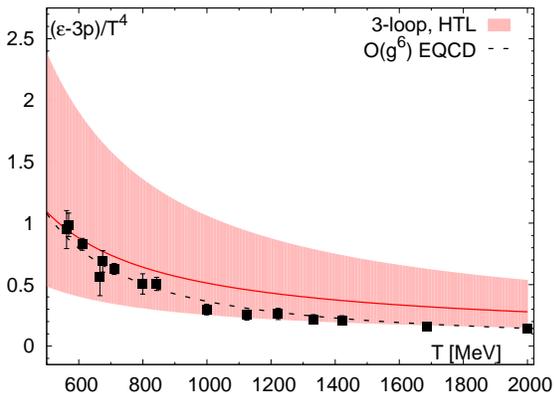}
\caption{
The comparison of the lattice results for the trace anomaly with
three-loop HTL perturbation theory shown as the line and the band.
The size of the band corresponds to the scale variation from $\mu=\pi T$ to
$4 \pi T$. Also shown as a deashed line is the EQCD result for the trace
anomaly (see text).
The lattice results for $N_{\tau}=4$ and $6$ for $T>600$ MeV have been 
scaled by $1.2$ and $1.4$, respectively.}
\label{fig:e-3p_comp}
\end{figure}

Now, that the cutoff dependence of the pressure is understood we can proceed 
with the continuum extrapolations. 
As discussed above at high temperatures the dominant cutoff dependence of the 
pressure is given by the cutoff dependence of the ideal quark gas, and 
therefore, for improved staggered actions like HISQ it is expected to
scale like $1/N_{\tau}^4$. 
This expectation is confirmed by the study of QNS at high temperatures with 
HISQ action \cite{Bazavov:2013uja,Ding:2015fca}.
On the other hand at low temperatures the dominant cutoff effects are due 
to taste-symmetry breaking of staggered fermions and scale like 
$a^2 \sim 1/N_{\tau}^2$. 
This is also confirmed by lattice calculations
\cite{Bazavov:2012jq}.
We find that the cutoff dependence of the pressure is incompatible with 
$1/N_{\tau}^2$ behavior for $T>400$ MeV. 
Similar findings have been obtained for 
QNS \cite{Bazavov:2013uja,Ding:2015fca}.
On the other hand, for $T<200$ MeV we find that $1/N_{\tau}^4$ behavior 
of the cutoff effects is incompatible with the data. 
Therefore, we will assume that the cutoff effects are proportional to 
$1/N_{\tau}^4$, when performing continuum extrapolations for $T>400$ MeV.
In the intermediate temperature region,
$200~{\rm MeV} < T < 400~{\rm MeV}$ 
the cutoff effects should be proportional to some combination
of $1/N_{\tau}^2$ and $1/N_{\tau}^4$. Thus, one should in principle
fit the data by $a/N_{\tau}^2+b/N_{\tau}^4$ form to obtain
the continuum limit. But because we have only four $N_{\tau}$ values
and the errors of the $N_{\tau}=10$ and $N_{\tau}=12$ data are large
the continuum result obtained from such fits has large statistical error.
It turns out, however, that in this intermediate temperature region
it is possible to fit the cutoff dependence of the pressure with
$1/N_{\tau}^2$ form as well as with $1/N_{\tau}^4$ form and obtain
$\chi^2/{\rm df} \sim 1$. The $1/N_{\tau}^2$ fits give higher values of the pressure
than the $1/N_{\tau}^4$ fits, though the results from both fits overlap within
the error bands. The difference between the central values of
the $1/N_{\tau}^2$ and $1/N_{\tau}^4$ fits could be considered as
a measure of the systematic error of the continuum extrapolation.
This difference turns out to be about the same as the statistical
errors of the $1/N_{\tau}^4$ extrapolations. Therefore, we estimate
the total error of the continuum pressure for $200~{\rm MeV} < T < 400~{\rm MeV}$ 
by doubling the statistical error of the $1/N_{\tau}^4$ fit. Alternatively
we could add the systematic error estimated as described above with
the statistical error in quadrature to obtain the total error. 
The corresponding errors will be smaller. We prefer to be more conservative.
Similar analysis as above has been performed to obtain
the continuum extrapolated result for the entropy density.

In the temperature interval $200$ MeV $<T< 660$ MeV we have four lattice 
spacings to perform continuum extrapolations. As the result the continuum 
extrapolations are most reliable in this temperature interval. 
In the temperature range $660$ MeV $<T< 800$ MeV we have three lattice 
spacings, so controlled continuum extrapolation is still possible. 
For $T>800$ MeV, we can only provide a continuum estimate for the pressure. 
For $T<1000$ MeV we do this by performing $const+1/N_{\tau}^4$ fit of the 
$N_{\tau}=6$ and $8$ data for the pressure. 
Interestingly, it turns out that the coefficient of $1/N_{\tau}^4$ obtained 
when using only $N_{\tau}=6$ and $8$ data, and when using $N_{\tau}=6,~8$ 
and $10$ data is the same within errors for $T\simeq 800$ MeV. 
So perhaps this extrapolation is not totally out of control.
Finally, to obtain the continuum estimate for $T>1000$ MeV we assume that 
the coefficient of the $1/N_{\tau}^4$ term is the same as obtained
from the fit of the $N_{\tau}=6$ and $8$ data at $T=995$ MeV and correct 
the $N_{\tau}=6$ pressure by $1/6^4$ times this coefficient. 
The continuum results for the pressure obtained using this procedure are 
shown in Fig. \ref{fig:p_cont} and compared to the corrected results for 
$N_{\tau}=6,~8,~10$ and $12$. 
The continuum result for the pressure agrees with the corrected results. 
This serves as an important cross-check for our continuum extrapolations 
for $T<1330$ MeV.
\begin{figure*}
\includegraphics[width=8cm]{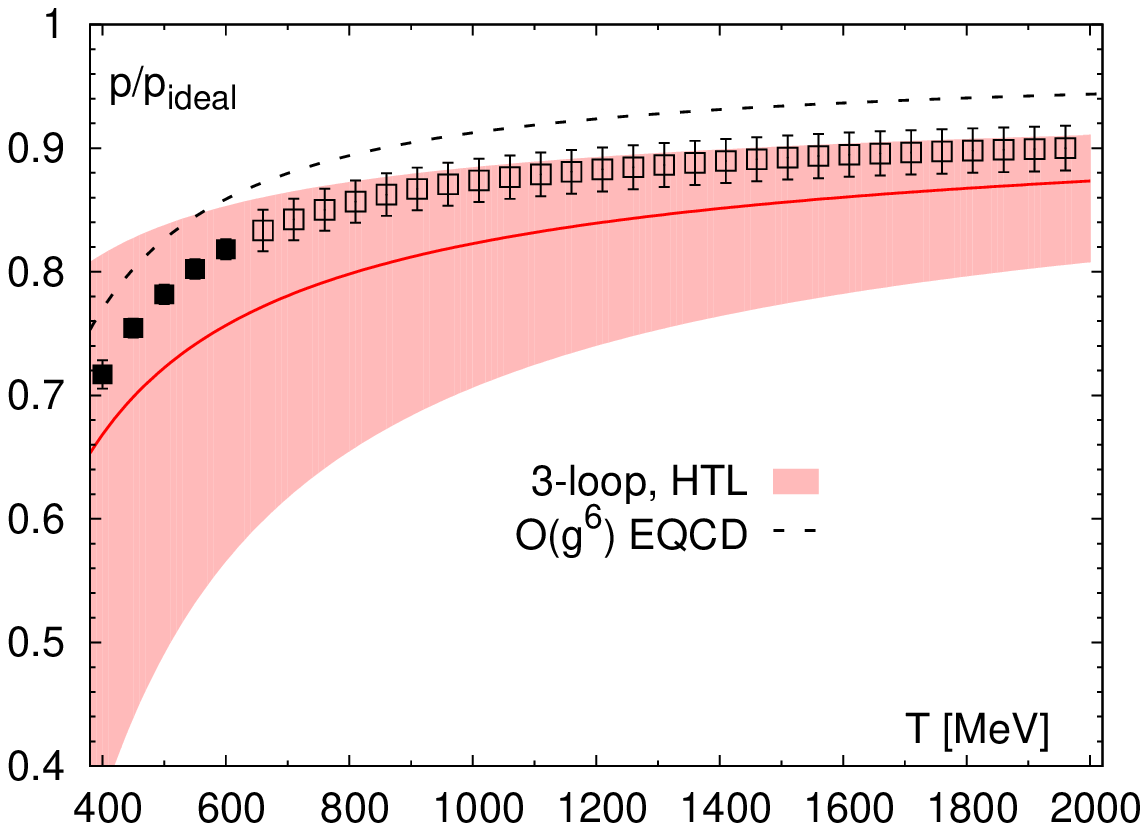}
\includegraphics[width=8cm]{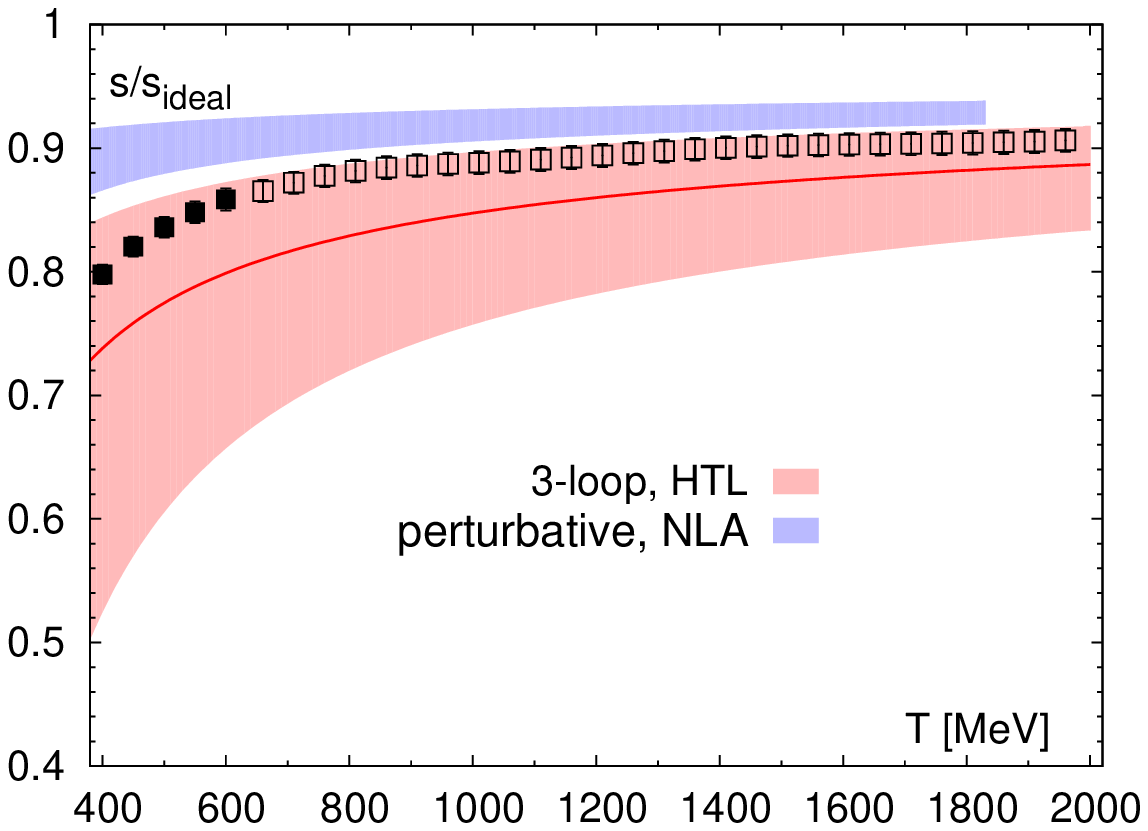}
\caption{The pressure (left) and the entropy density (right) in the high
temperature region compared with the weak-coupling calculations.
The filled symbols correspond to the continuum results obtained
from lattice calculations on $N_{\tau}=6,~8,~10$ and $12$ lattices.
The open symbols correspond to continuum estimate (see text).
The errors of the continuum estimate have been enlarged by factor
two to indicate additional systematic errors that might be present.
The red line and the band correspond to the three-loop HTL perturbation theory \cite{Haque:2014rua},
the blue band corresponds to the resummed calculation in next-to-leading
log approximation (NLA) \cite{Rebhan:2003fj}.
The width of the bands correspond to the scale variation from $\mu=\pi T$
to $4 \pi T$.
Also shown is the weak-coupling result obtained in EQCD \cite{Laine:2006cp}.}
\label{fig:eos_high}.
\end{figure*}

For comparison of the lattice calculation of the EoS with the weak-coupling 
results it is important to have an alternative method to obtain continuum 
results for $T>800$ MeV and also extend the calculation to higher 
temperatures. 
In order to do this we consider again the trace anomaly. 
As discussed in Section \ref{secTmumu} for $N_{\tau}\ge 8$ and $T>300$ MeV 
we do not see any cutoff dependence of the trace anomaly. 
This also means that given the statistical errors the $N_{\tau}=8$ results 
for the trace anomaly can be considered as the continuum ones. 
Therefore, we can perform a combined interpolation of the numerical results 
for the trace anomaly obtained with $N_{\tau}=8,~10$ and $12$ in the 
temperature interval $300$ MeV $<T< 1000$ MeV, providing a continuum estimate. 
The $N_{\tau}=4$ and $6$ results for the trace anomaly lie below this 
continuum estimate. 
However, if we re-scale the $N_{\tau}=4$ and $6$ results on the trace anomaly 
by factors $1.2$ and $1.4$, respectively, they agree with the above continuum 
estimate for $800$ MeV $<T<1000$ MeV within errors.
This is demonstrated in Fig. \ref{fig:e-3p_corr}.
Therefore, to obtain a continuum estimate for the trace anomaly beyond 
$T=1000$ MeV we re-scale the $N_{\tau}=4$ and $N_{\tau}=6$ data for 
$T>1000$ MeV with the above factors. 
Here we tacitly assume that
the cutoff dependence of the trace anomaly is temperature independent.
This assumption, however, is quite reasonable since the cutoff dependence
at high temperatures should be described by weak-coupling expansion and thus
is proportional to $a^2=1/(N_{\tau} T)^2$ times the coupling constant to some
power. Since the coupling constant depends on the temperature scale logarithmically
in a limited temperature interval the cutoff effects should be approximately
temperature independent. Our study of the $N_{\tau}$ dependence of
the pressure for $T>400$ MeV confirms this expectation. The cutoff
dependence of the quark number susceptibilities \cite{Bazavov:2013uja,Ding:2015fca}
and the free energy of the static quark \cite{Bazavov:2016uvm} also support this
assumption. Therefore, we perform a spline interpolation of the combined $N_{\tau}=12,~10,~8,~6$ 
and $4$ data in the temperature interval $400$ MeV $<T<2000$ MeV.
Because we corrected the trace anomaly
obtained on $N_{\tau}=4$ and $N_{\tau}=6$ lattices we
assign an additional systematic
error of $20\%$  and $40\%$ to the corresponding data points before the interpolation, i.e. the size
of the systematic errors that we assume is the same as the magnitude of the correction.
Using this interpolation we calculate the integral of the trace anomaly from 
$T=660$ MeV to $2000$ MeV, which together with the continuum result for the 
pressure at $660$ MeV obtained above gives us the continuum pressure estimate 
that extends to temperatures as high as $2000$ MeV. 
From the pressure we can also calculate the entropy density. 
These calculations will be used in the next section for the comparison with 
the weak-coupling results.
We also compared this continuum estimate of the pressure with the one 
discussed before.
For $T<1330$ MeV we find excellent agreement between the two continuum 
estimates. 

We note that our continuum result for $T=500$ MeV is one and a half sigma 
higher than the continuum result of Ref. \cite{Borsanyi:2013bia}, while our 
continuum estimate for higher temperatures is $5-7\%$ higher than the 
continuum estimate of Ref. \cite{Borsanyi:2010cj}. Our continuum result
for the pressure for $T<400$ MeV agrees very well with the HotQCD result
\cite{Bazavov:2014pvz} but has considerably smaller errors.

\section{Equation of state at high temperatures and comparison with 
weak-coupling calculations}

In this section we compare the lattice results on the EoS with the weak-coupling 
calculations.
We start our discussion with the trace anomaly.
In Fig. \ref{fig:e-3p_comp} we  compare our lattice results for the trace 
anomaly obtained with $N_{\tau}=8,~10$ and $12$ as well as the corrected 
results for $N_{\tau}=4$ and $6$ (see previous section) with the results of
three-loop HTL perturbation 
theory \cite{Haque:2014rua}.
We see good agreement between the lattice results and the results obtained in three-loop 
HTL perturbation theory, although the error band of the latter is still 
quite large. 
The lattice results on the trace anomaly agree very well with the 
weak-coupling calculations based on dimensionally reduced effective field 
theory, the electrostatic QCD (EQCD) \cite{Laine:2006cp}.

Next we compare the high temperature lattice results for the pressure and 
the entropy density  with the three-loop HTL perturbation theory 
\cite{Haque:2014rua} and the results obtained using EQCD \cite{Laine:2006cp}. 
The comparison is shown in Fig. \ref{fig:eos_high}. 
For $T<660$ MeV we use
the continuum extrapolated lattice results obtained from the calculations on 
$N_{\tau}=6,~8,~10$ and $12$ lattices. For higher temperatures, we use the continuum
estimate based on the trace anomaly calculated on 
the coarsest $N_{\tau}=4$ lattice.
As discussed in the previous section this continuum
estimate is validated by direct continuum extrapolation for $T<1000$ MeV, but
at higher temperatures it relies on the temperature independence of the cutoff
effects. Therefore, in Fig. \ref{fig:eos_high} we show the continuum estimate
as open symbols.
We see that the EQCD calculations are higher than 
our lattice results. Our lattice results lie above the central value of the three-loop HTL 
perturbative result by one sigma. 
However, the lattice data are fully contained within the 
uncertainty of the three loop HTL result.
In the considered temperature range the uncertainty of the lattice results is 
significantly smaller than the uncertainty of the three-loop HTL result. 
For the entropy density we also compare our lattice results with
the resummed perturbative calculations in next-to-leading
log approximation (NLA) \cite{Rebhan:2003fj}.
This comparison is shown in Fig. \ref{fig:eos_high} (right).
The NLA calculation leads to higher entropy density than the lattice result,
although overlaps within the uncertainty with latter for $T>1300$ MeV.
The NLA calculation is based on the $\Phi$ derivable approach 
\cite{Blaizot:1999ip,Blaizot:1999ap,Blaizot:2000fc}. In this approach
one calculates the derivatives of the pressure, which leads to cancellation
of many higher order diagrams. As the result one obtains relatively simple expressions
for the entropy density \cite{Blaizot:2000fc} or the quark number susceptibility
\cite{Blaizot:2001vr}. The calculation of the pressure in this approach, however,
is difficult.

It is clear that our lattice results are sufficiently precise to test the 
various weak-coupling approaches and it would be desirable to further reduce 
the uncertainty of the weak-coupling approaches to see if the thermodynamics 
of the quark gluon plasma can be indeed understood using the weak-coupling 
expansion in the considered temperature range.

\section{Conclusions}

We extended the previous calculation of the EoS with HISQ action to 
higher temperatures.
First, we extended the calculation of the trace anomaly to higher 
temperatures using lattice simulations at larger quark mass, $m_l=m_s/5$. 
We showed that the quark mass dependence of the trace anomaly is negligible 
for $T>400$\,MeV given the statistical error.
Then using the results on the trace anomaly and the integral method we 
calculated the pressure for $N_{\tau}=6,~8,~10$ and $12$. 
We studied the cutoff ($N_{\tau}$) dependence of the pressure and performed 
the continuum extrapolation in the high temperature limit. 
We pointed out that the cutoff dependence of the pressure is dominated by 
the quark contribution and the cutoff dependence of this contribution is 
very similar to the cutoff dependence of QNS at high temperatures. 
We also showed that using the known cutoff dependence of QNS it is possible 
to correct for the cutoff effects in the pressure at fixed $N_{\tau}$. 
The corrected results for the pressure calculated with HISQ action and 
p4 action for different $N_{\tau}$ agree within errors and also agree with 
the continuum result. 
Thus, we achieved a controlled continuum extrapolation
of the pressure at 
high temperatures.  
Finally, using $N_{\tau}=4$ and $6$ results on the trace anomaly we provided 
a continuum estimate for the pressure that extends to temperatures as high 
as $T=2000$\,MeV.
We compared this continuum estimate with the weak-coupling calculations and 
found a reasonably good agreement between the lattice and the weak-coupling 
results.

\section*{Acknowledgements}
The simulations have been carried out on the computing facilities of 
the Computational Center for Particle and Astrophysics (C2PAP), SuperMUC and 
NERSC. We used the publicly available MILC code to perform the numerical
simulations \cite{milc}.
The data analysis was performed using the R statistical package 
\cite{Rpackage}. We thank F. Karsch for providing the numerical values
of the free quark pressure for finite $N_{\tau}$.
We also thank M. Strickland and N. Haque for sending the 3-loop HTL results
for the EoS.
This work has been supported in part by the U.S Department of Energy through grant contract No. DE-SC0012704.
J. H. Weber acknowledges the support by the Bundesministerium f\"ur Bildung und 
Forschung (BMBF) under grant ``Verbundprojekt 05P2015 - ALICE at High Rate 
(BMBF-FSP 202) GEM-TPC Upgrade and Field theory based investigations of ALICE 
physics'' under grant No. 05P15WOCA1.

\appendix
\section{Zero temperature calculations}\label{appA}

For $\beta=7.03$ and $7.825$ we generated a single stream of MC evolution. 
For the highest three $\beta$ values we generated three streams of MC 
evolution called a, b and c; each of these streams corresponds to a single 
value of topological charge $Q$. 
The lengths of these streams for every value of $\beta$ are  $1389$, $1269$ and 
$1269$, respectively.
The values of the plaquette, rectangles, light and strange quark condensates 
are given in Tab. \ref{tab:Qdep} together with the values of $Q$. 
The values of rectangles and plaquettes are the same within errors for 
the streams with different $Q$. 
For the quark condensate we see small, but in some cases statistically
significant differences. 
The difference in the value of the light quark condensate is between 
$1\%$ and $2\%$, while for the strange quark condensates it is $<0.5\%$.
For $\beta=8.4$ there are differences in the values of the light quark condensate
also for the streams that belong to the same topological sector, which appear
to be statistically significant. This is most likely due to the fact that each
of the streams is relatively short.
These small differences, are taken into account in
the calculations as additional systematic errors. However, the
additional systematic effects
are largely irrelevant for the calculation of the 
trace anomaly, since the highest three beta values correspond to temperatures 
$T>400$ MeV, where the contribution of quark condensates is very small.

\begin{table*}
\begin{tabular}{ccccccc}
$\beta$ & plaq.  & $\langle \bar \psi \psi \rangle_l$ & $\langle \bar \psi \psi \rangle_s$ & rect. & Q & stream\\
\hline
       & 0.6641244(21) &  0.0026400(66)  & 0.0117787(51) &   0.4607658(29) &  2  & a \\
8.0    & 0.6641256(15) &  0.0026061(122) & 0.0117719(76) &   0.4607666(23) &  1  & b \\
       & 0.6641257(24) &  0.0025923(87)  & 0.0117889(59) &   0.4607666(35) &  0  & c \\
\hline
       & 0.6738855(16) &  0.00207849(52) & 0.0095507(55) &   0.4744956(24) &  2  & a \\
8.2    & 0.6738854(12) &  0.00199916(69) & 0.0095271(60) &   0.4744943(17) &  0  & b \\
       & 0.6738865(12) &  0.00201003(95) & 0.0095399(75) &   0.4744971(18) &  0  & c \\
\hline
       & 0.6830217(14) &  0.00171386(47) & 0.0078134(48) &   0.4874515(22) &  2  & a \\
8.4    & 0.6830200(17) &  0.00158675(57) & 0.0077629(71) &   0.4874514(28) &  0  & b \\
       & 0.6830187(12) &  0.00161808(63) & 0.0077963(54) &   0.4874474(18) &  0  & c \\
\hline
\end{tabular}
\caption{The values of plaquette, rectangle, light and strange quark 
condensates at $T=0$ for different topological sectors $Q=0,~1,~2$.}
\label{tab:Qdep}.
\end{table*}

\begin{figure}
\includegraphics[width=7cm]{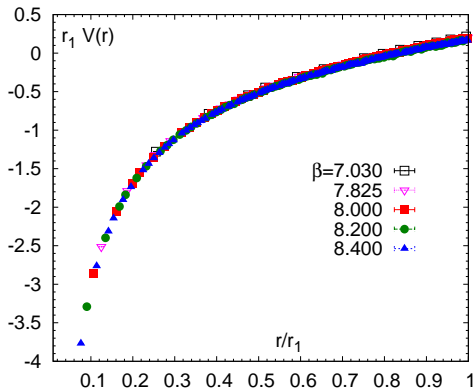}
\caption{The potentials calculated for $m_l=m_s/5$ at $\beta=7.030,~7.825,~8.00,~8.20$ and $8.40$.}
\label{fig:pot}
\end{figure}

To determine the lattice spacing we calculated the static quark anti-quark 
potential. 
We followed the same procedure as in Ref. \cite{Bazavov:2014pvz}, in 
particular the same fit ranges in time were used.
For the highest three $\beta$ values we used a fit range in $t/a$, which is 
about the same as in Ref. \cite{Bazavov:2014pvz} for $\beta=7.825$. 
Our results for the potential are shown in Fig. \ref{fig:pot}. 
We calculated the static quark anti-quark potential for different topological 
sectors and did not see any dependence on the topological charge within the 
statistical errors.

\begin{figure}
\includegraphics[width=7cm]{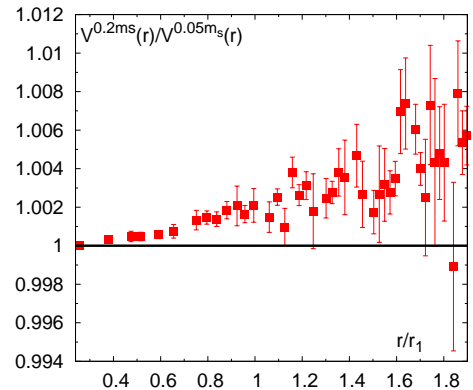}
\caption{The ratio of the potentials calculated for $m_l=m_s/20$ and $m_l=m_s/5$ for $\beta=7.03$.} 
\label{fig:pot_rat}
\end{figure}

It is interesting to compare the potential calculated for $m_l=m_s/5$ and  
$m_s/20$ at the same value of $\beta$. 
Such a comparison is shown in Fig. \ref{fig:pot_rat} for $\beta=7.03$. 
As one can see from the figure the quark mass effects are very small 
for $r<r_1$. 
They are the smallest at the shortest distance and gradually increase with 
increasing $r$. 
However, even for $r=r_1$ the effects are smaller than $0.2\%$ and 
for $r<0.8 r_1$ are smaller than $0.1\%$. 
We get very similar results for $\beta=7.825$.

We also find that the difference between the potential calculated for $m_s/5$ 
and $m_s/20$ in units of $r_1$ can be parametrized as 

\begin{equation}
\Delta V(r)=V^{m_s/5}(r)-V^{m_s/20}(r)=b (r/r_1)^2.
\end{equation}

We get $b=0.00577(19)$ for $\beta=7.03$ and $b=0.00774(51)$ for $\beta=7.825$.
>From these we can estimate that $r^2 V''(r)$ is changed by about $1.4\%$ 
around $r=r_1$, and by about $0.2\%$ or less around $r=r_2$, when changing 
$m_l$ from $m_s/20$ to $m_s/5$. 
Therefore, we expect shifts in the values of $r_1$ and $r_2$ with changing 
quark masses, which are similar in magnitude.

\begin{figure}
\includegraphics[width=7cm]{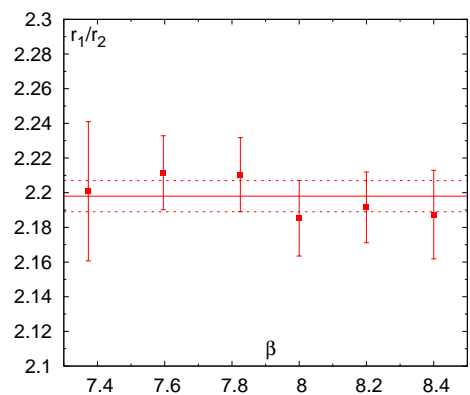}
\caption{The ratio of the ratio of scales $r_1$ and $r_2$ as function of 
$\beta$. 
Also shown is the fit (solid line) and its uncertainty (dashed lines).}
\label{fig:rat12}
\end{figure}

The $r_1$ scale obtained from the potential at $\beta=7.03,~7.825,~8.00,~8.20$ 
and $8.40$ is given in Tab. \ref{tab:r12}. 
For $\beta \ge 7.825$ we also calculated the $r_2$ scale. 
Moreover, this scale was calculated for $m_l=m_s/20$ using the data on the 
potential from \cite{Bazavov:2014pvz}. 
The results are given in Tab. \ref{tab:r12}. 
We see that the $r_1$ scale is about $1\%$ smaller for $m_l=m_s/5$ than for 
$m_s/20$. 
This difference is consistent with the above expectations and statistically it 
is not very significant. 
The value of $r_2$ at $\beta=7.825$ is $0.3\%$ smaller for $m_l=m_s/5$ than 
for $m_l=m_s/20$. 
Again this difference is statistically not significant.
Since the $r_2$ scale shows smaller quark mass dependence we could use it to 
extend the scale setting procedure of Ref. \cite{Bazavov:2014pvz} to higher 
$\beta$, namely up to $\beta=8.40$.
To do this we first consider the ratio $r_2/r_1$, which is shown in 
Fig. \ref{fig:rat12}. 
We do not see any $\beta$ dependence of this ratio within errors.
\begin{table}
\begin{tabular}{ccll}
\hline
$\beta$ & $m_l/m_s$ & $r_1$ & $r_2$ \\
\hline
8.400  & 1/5 &  12.560(130)  &  5.742(31) \\
8.200  & 1/5 &  10.653(60)   &  4.861(36) \\
8.000  & 1/5 &   8.905(60)   &  4.075(30) \\
7.825  & 1/5 &   7.570(104)  &  3.469(18) \\
7.825  & 1/20 &   7.690(58)   &  3.479(21) \\
7.596  & 1/20 &   6.336(56)   &  2.865(11) \\
7.373  & 1/20 &   5.172(34)   &  2.350(40) \\
7.030  & 1/5 &   3.737(13)   &  -         \\
7.030  & 1/20 &   3.763(13)   &  -         \\
\hline
\end{tabular}
\caption{
The value of the scale parameters for different $\beta$ values and quark 
masses used in this study.}
\label{tab:r12}
\end{table}
Fitting $\beta>7.825$ we get $2.188(12)$, while the fit with 
$\beta \ge 7.825$ we obtain $2.210(14)$. 
Finally the fits for all $\beta$ in the interval $[7.373:8.400]$ we get 

\begin{equation}
\left(\frac{r_1}{r_2}\right)_{av}=2.198 \pm 0.009, 
\label{rat12}
\end{equation}

\noindent which agrees with the above values within errors.
Since $r_2$ is essentially mass independent and more accurately determined 
than $r_1$ for the highest $\beta$ values 
we will use it for the scale setting. 
We combine the results of Ref. \cite{Bazavov:2014pvz} together with 
$(r_1/r_2)_{av}\cdot r_2$ for $\beta=7.596,~7.825,~8.000,~8.200$ and $8.400$ 
from Tab. \ref{tab:r12} to obtain the lattice spacing in units of $r_1$ in 
the $\beta$ region that extends to $\beta=8.400$.
As in  Ref. \cite{Bazavov:2014pvz} we fit $a/r_1$ with an Allton-type form \cite{Allton:1996dn}:

\begin{align}
\displaystyle
\frac{a}{r_1}
&=
\frac{c_0 f(\beta)+c_2 (10/\beta) f^3(\beta)}{
1+d_2 (10/\beta) f^2(\beta)}\; ,\label{ar1_fit} \\[3mm]
\displaystyle
f(\beta)
&=
\left( \frac{10 b_0}{\beta} \right)^{-\displaystyle\frac{b_1}{2 b_0^2}} \exp\left(-\displaystyle\frac{\beta}{20 b_0}\right)\; .
\label{fbeta}
\end{align}

\noindent Here $b_0$ and $b_1$ are the well-known coefficients of the two-loop beta 
function, which for the three-flavor case read $b_0=9/(16 \pi^2)$, 
$b_1=1/(4 \pi^4)$. 
Fitting the combined data set for the coefficients $c_0,~c_2$ and $d_2$ 
we get:

\begin{align}
c_0&=43.12 \pm 0.18\; ,\\
c_2&=347008 \pm 32131\; ,\\
d_2&=5584 \pm 599\;,\\
\chi^2/{\rm df}&=0.25.
\end{align}

\begin{figure}
\includegraphics[width=7cm]{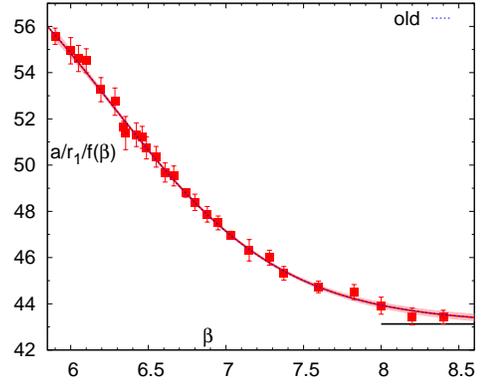}
\caption{The lattice spacing in units of $r_1$ as function of $\beta$. For
better visibility we divided $a/r_1$ by the 2-loop beta function given by Eq. (\ref{fbeta}).
The horizontal line shows the asymptotic 2-loop result. 
Also shown as a dashed line is the old parametrization of $a/r_1$ from 
Ref. \cite{Bazavov:2014pvz}.}
\label{fig:r1fit}
\end{figure}

The above errors have been estimated by bootstrap method and they are 
smaller than those in \cite{Bazavov:2014pvz}, in particular the error on 
$c_0$ is reduced from $0.3$ to $0.18$. 
This fit is shown in Fig. \ref{fig:r1fit} with the band indicating its
uncertainty. 
The difference between this parametrization of $a/r_1$ and the one in 
Ref. \cite{Bazavov:2014pvz} is less than $0.3\%$ in the entire range of 
$\beta$. 
It is interesting to note that for the highest $\beta$ value the deviation 
from the asymptotic 2-loop result is only one sigma. 
>From this fit we can determine the smoothed value of $r_1/a$ for each value 
of $\beta$ and thus the temperature scale.

\begin{figure}
\includegraphics[width=7cm]{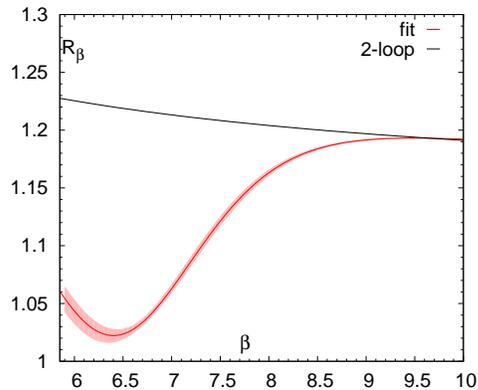}
\caption{The logarithmic derivative of the non-perturbative beta function.}
\label{fig:Rbeta}
\end{figure}

For the calculation of the EoS we also need the non-perturbative beta function
defined as 

\begin{equation}
R_{\beta}=-a \frac{d \beta}{da}= \frac{r_1}{a} \left( \frac{d (r_1/a)}{d \beta} \right)^{-1} \; .
\label{rbeta}
\end{equation}

This is shown in Fig. \ref{fig:Rbeta}.
We also calculated the mass of the unmixed $\eta_{ss}$ meson for $\beta=8.0$
and obtained $a m_{\eta_{ss}}=0.12282(40)$. This value agrees well with 
determination of $m_s^{LCP}(\beta)$ given in Ref. \cite{Bazavov:2014pvz}. 
For $\beta=8.2$ and $8.4$ the lattices are too small to determine 
$a m_{\eta_{ss}}$ reliably.

\section{The hadron resonance gas and cutoff effects at low temperatures}\label{appB}

In the HRG model the partition function of strongly interacting matter 
at low temperatures is given by the partition function 
of non-interacting hadrons and resonances

\begin{align}
p^{HRG}/T^4
&=\frac{1}{VT^3}\sum_{i\in\;mesons}\hspace{-3mm} 
\ln{\cal Z}^{M}(m_i,T,V)
\nonumber \\
&+\frac{1}{VT^3}
\sum_{i\in\;baryons}\hspace{-3mm} \ln{\cal Z}^{B}(m_i,T,V)\; ,
\label{eZHRG}
\end{align}

\noindent where

\begin{equation}
\ln{\cal Z}^{M/B}(m_i,T,V)
=\mp \frac{V d_i}{{2\pi^2}} \int_0^\infty dk k^2
\ln(1\mp e^{-E_i/T}) \quad ,
\label{ZMB}
\end{equation}

\noindent with energies $E_i=\sqrt{k^2+m_i^2}$ and degeneracy factors $d_i$.
The superscripts $M$ and $B$ refer to mesons and baryons.
Usually the sum in the above equation contains all the meson and baryons from 
the Particle Data Group (PDG).
However, our information of the baryon spectrum may be incomplete. 
There are lots of baryon states predicted by the quark model 
(QM) \cite{Capstick:1986bm} as well as by lattice 
QCD \cite{Edwards:2012fx} that are not included in the PDG. 
These are the so-called missing states. 
It was shown that these missing states are important for QCD 
thermodynamics \cite{Majumder:2010ik,Bazavov:2014xya,Bazavov:2014yba,Bazavov:2017dus,Bazavov:2017tot}. 
Therefore, we included these missing states in the HRG model. 
We used the baryon spectrum from the quark model calculations of 
Refs. \cite{Loring:2001kx,Loring:2001ky}.
We call this model HRG-QM.
For the strange baryons we also used the spectrum from 
Ref. \cite{Capstick:1986bm} and found that this only results in very small 
differences relative to the above calculation.
The difference between the HRG-QM and the HRG model which includes only 
hadrons 
from PDG, and therefore is called HRG-PDG, is visible only for $T>150$\,MeV. 
At these temperatures, however, the HRG model itself may not be reliable.

The hadron resonance gas can be used as a tool to understand the cutoff 
effects in the EoS at low temperatures.
Lattice discretization errors will modify the hadron spectrum which then 
leads to the modification of the HRG.
This has been discussed in some details for p4 and asqtad 
actions \cite{Huovinen:2009yb}.
Below we will discuss the discretization effects in the hadron spectrum
for the HISQ action and their effect on thermodynamics of hadrons.

The staggered fermion formulation describes four flavors (tastes) of quarks in 
the continuum limit.
To describe a single quark flavor one takes the fourth root of the 
staggered fermion determinant in the path integral of QCD. 
This is the so-called rooting trick and amounts to averaging over the four 
staggered tastes for each physical flavor.
For the discussion of the cutoff effects on the hadron spectrum we first 
limit ourselves to the original four-flavor case. 
There are 16 pseudo-scalar (ps) mesons, which are the Goldstone bosons of 
the theory.
At non-zero lattice spacing only a $U(1)$ subgroup of the $SU(4)_A$ group 
is preserved, and there is only one Goldstone boson in the chiral limit. 
The other ps mesons have squared masses proportional to $a^2$, 
$\delta m_{{\rm ps}_i} e_i a^2$. 
The breaking of the full chiral symmetry to a $U(1)$ subgroup and the 
corresponding splitting of ps mesons is referred to as taste-symmetry 
breaking. 
It is the largest source of discretization errors in today's lattice 
calculations with staggered fermions. 
The size of taste-symmetry breaking, i.e. the value of coefficients $e_i$ can be 
reduced by using improved actions.
All improved staggered actions (p4, asqtad, stout and HISQ) reduce the size 
of taste-symmetry breaking to some degree,
The HISQ action has the smallest taste-symmetry breaking among the improved 
staggered fermion actions \cite{Bazavov:2011nk}.
The taste symmetry breaking effects are particularly large for the p4 action.

\begin{figure*}
\includegraphics[width=8cm]{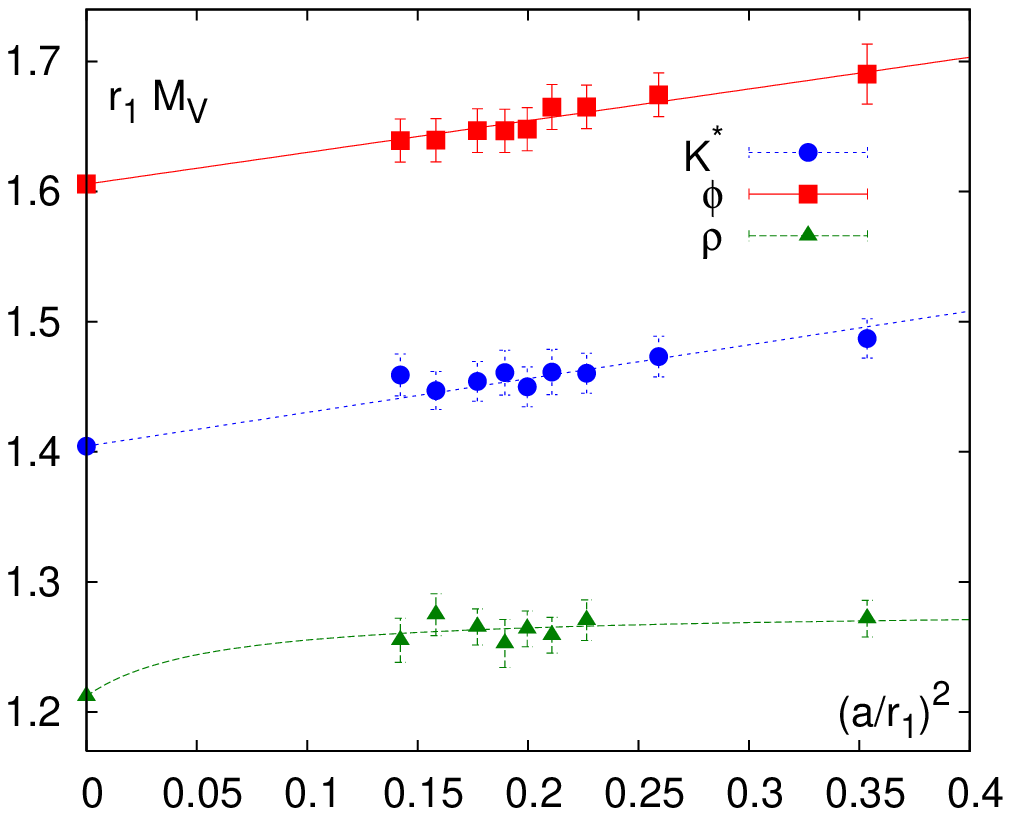}
\includegraphics[width=8cm]{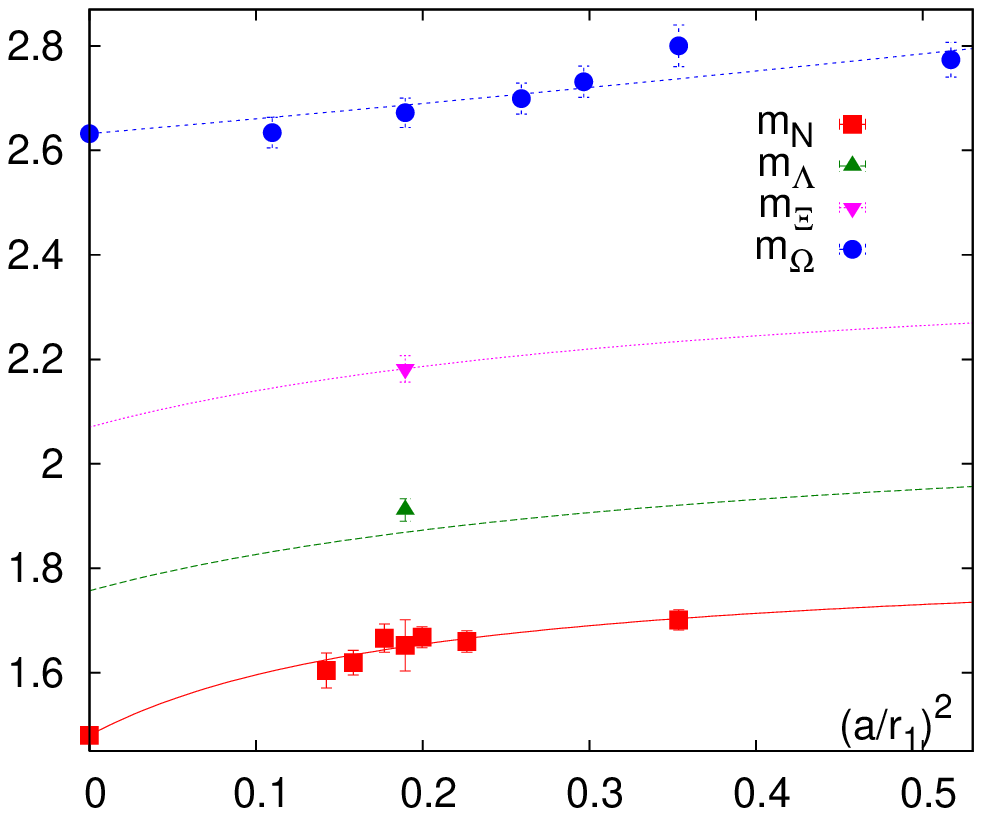}
\caption{The lattice spacing dependence of vector meson (left) and baryon 
masses (right).
The lines show fits using Eq. (\ref{mH}).}
\label{fig:mH}
\end{figure*}

Taste-symmetry breaking also causes non-degeneracy of vector mesons and baryons that 
belong to different tastes. 
However, the corresponding mass splittings are much smaller than in
the case of ps mesons. 
For the HISQ action they are of the size of statistical errors and 
therefore can be neglected in the following discussion. 
The dominant effects of taste-symmetry breaking in the vector meson and baryon sectors 
come from the fact that the calculations are effectively performed at larger 
value of the pion mass than the physical one if the lattice spacing is 
non-zero. 
Since hadronic quantities like hadron masses and decay constants decrease 
with decreasing pion masses, we expect that the continuum limit for these 
quantities is approached from above. 
The masses of the vector mesons, nucleons and $\Omega$ baryons have been 
calculated with HISQ action for different lattice 
spacings \cite{Bazavov:2011nk,Bazavov:2014pvz}.
We complement these studies by also calculating the masses of octet baryons with 
strangeness $S=-1$ and $S=-2$ for $\beta=10/g^2=6.515$ corresponding to lattice 
spacing $a=0.135$\,fm. 
In Fig. \ref{fig:mH} we show the vector meson and baryon masses as function of 
the lattice spacing. 
We see that following our expectations the hadron masses approach their
continuum limit from above. 
We fit the $a$-dependence of the hadron masses by the form
\begin{equation}
r_1 m_H=(r_1 m_H)^{cont}+\frac{b_H (a/r_1)^2}{1+c_H (a/r_1)^2}.
\label{mH}
\end{equation}
The values of $b_H$ and $c_H$ are given in Table \ref{tab:mH}. 
The resulting fits are also shown in Fig. \ref{fig:mH} as lines and describe 
the data fairly well. 
For $S=-1$ and $S=-2$ baryons we could not perform the above fits. We model
their lattice spacing dependence using Eq. (\ref{mH}) with coefficients $b_H$
and $c_H$ obtained for the nucleon and divided by two and three, respectively.
This seems to capture the cutoff effects in $S=-1$ and $S=-2$, see Fig. \ref{fig:mH}.

To reduce cutoff effects in the thermodynamic quantities it has been 
suggested to use the kaon decay constant $f_K$ to set the lattice spacing. 
Since $f_K$ shows an $a$-dependence that is similar to that of the hadron 
masses the ratios $m_H/f_K$ are expected to have much milder $a$-dependence. 
As the consequence thermodynamic quantities will also have smaller cutoff dependence if $f_K$ is used to set the scale. 
We checked that the $a$-dependence almost entirely disappears for $K^*$ and $\phi$ mesons, as well as for the $\Omega$ 
baryon if $f_K$ is used to 
set the lattice spacing.
However, for the nucleon and other baryons this is not the case. 
Furthermore, the large taste-symmetry breaking in the ps meson sector cannot be 
compensated by changing scale from $r_1$ to $f_K$. 
\begin{table}
\begin{tabular}{|lccccc|}
\hline
       & $\rho,~\omega$  &  $K^*$   & $\phi$  & $N$     & $\Omega$ \\ 
\hline
$b_H$  & 1.2138          & 0.259522 & 0.24377 & 1.85148 & 0.306749 \\
$c_H$  & 18.1236         & 0        & 0       & 5.42284 & 0        \\
\hline
\end{tabular}
\caption{The values of the coefficients $b_H$ and $c_H$ entering 
Eq. \eqref{mH} for different hadrons.}
\label{tab:mH}
\end{table}

To take into account the effects taste-symmetry breaking in the ps meson 
sector the contributions of pions, kaons and eta mesons are calculated 
as \cite{Huovinen:2009yb}
\begin{equation}
p^{\pi,K,\eta}/T^4=\frac{1}{16} \frac{1}{VT^3} \sum_{i=0}^7 d_{{\rm ps}_i} \ln{\cal Z}^{M}(m_{{\rm ps}_i},T,V),
\label{ps_contr}
\end{equation}
\noindent where $m_{{\rm ps}_i}^2=m_{\pi,K,\eta}^2+\delta m_{{\rm ps}_i}^2$. 
The quadratic pseudo-scalar meson splittings have been calculated in 
Ref. \cite{Bazavov:2011nk}. 
The dependence of these splittings can be fitted well by the form
\begin{equation}
r_1 \delta m_{{\rm ps}_i}^2=\frac{e_i (a/r_1)^2}{1+g_i (a/r_1)^2}.
\label{mps}
\end{equation}
The value of the coefficients $e_i$ an $g_i$ together with the
degeneracy factors $d_{{\rm ps}_i}$ are given in Table \ref{tab:ps}.
\begin{table*}
\begin{tabular}{|lcccccccc|}
\hline
  i            & 0 & 1  & 2 & 3 & 4 & 5 & 6 & 7 \\
\hline
$d_{{\rm ps}_i}$ & 1 & 1        & 3        & 3        & 3       &  3       & 1        & 1       \\
$e_i$          & 0 & 8.34627  & 8.17699  & 14.6245  & 16.0450 & 21.1623  & 23.0067  & 30.8425 \\
$g_i$          & 0 & -4.83538 & -6.09594 & -6.72714 & -5.2249 & -6.47337 & -5.49115 & -3.64465 \\
\hline
\end{tabular}
\caption{The values of the coefficients $e_i$ and $g_i$ entering 
Eq. \eqref{mps} for different tastes of ps mesons as well as the degeneracy factors 
$d_{{\rm ps}_i}$.}
\label{tab:ps}
\end{table*}

The contribution of the ground state vector mesons can be evaluated at 
non-zero lattice spacing using Eqs. \eqref{ZMB} and \eqref{mH} and the 
corresponding values of $b_H$ and $c_H$ from Table \ref{tab:mH}. 
The $a$-dependence of the octet baryon masses, as well as of $\Omega$ mass 
is fixed through Eq. \eqref{mH} and the values of the coefficients are given 
in Table \ref{tab:mH}. 
To completely specify the contribution of the ground state baryons to the 
partition function we assume that the masses of the decuplet baryons for 
$S=0,-1$ and $-2$ have the same $a$-dependence as their octet partners. 
Thus the contribution of all ground state hadrons at non-zero $a$ is now 
fixed.

\begin{figure*}
\includegraphics[width=7cm]{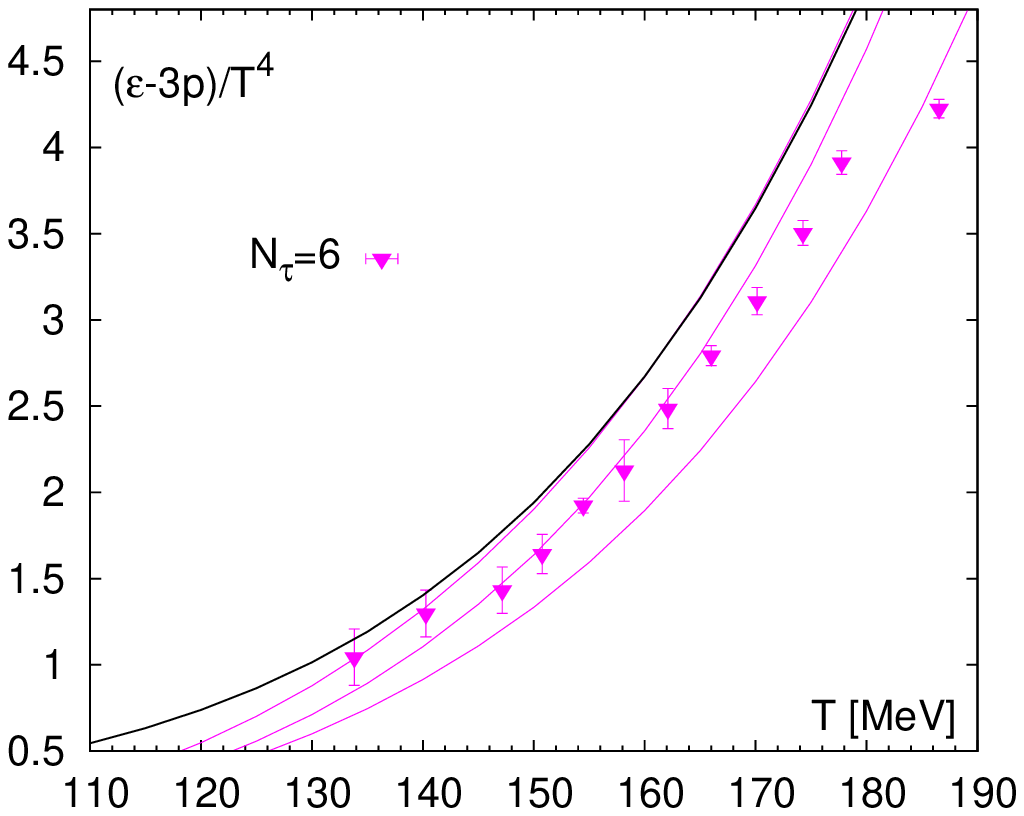}
\includegraphics[width=7cm]{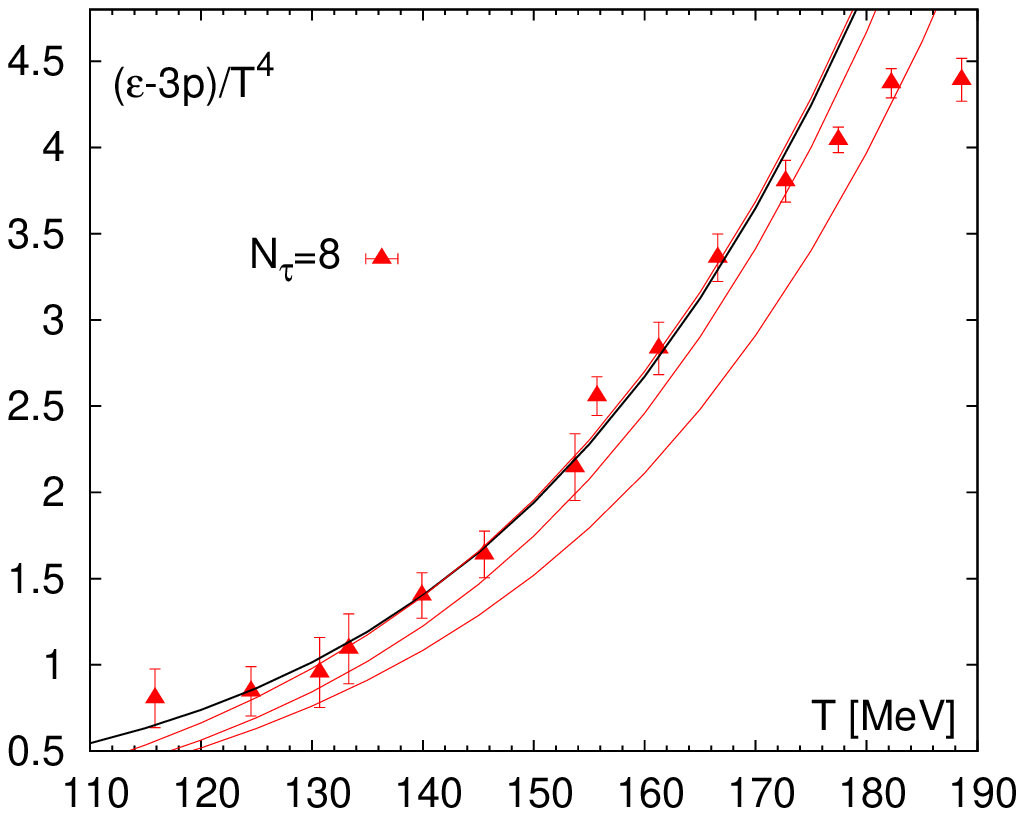}
\includegraphics[width=7cm]{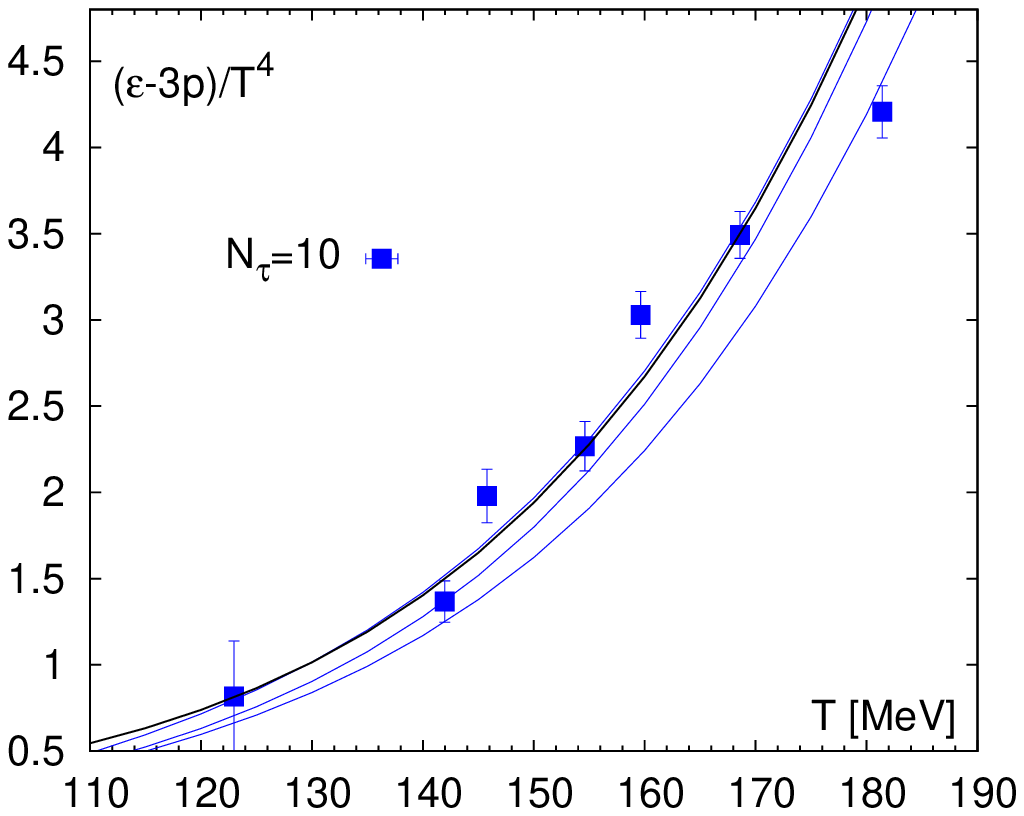}
\includegraphics[width=7cm]{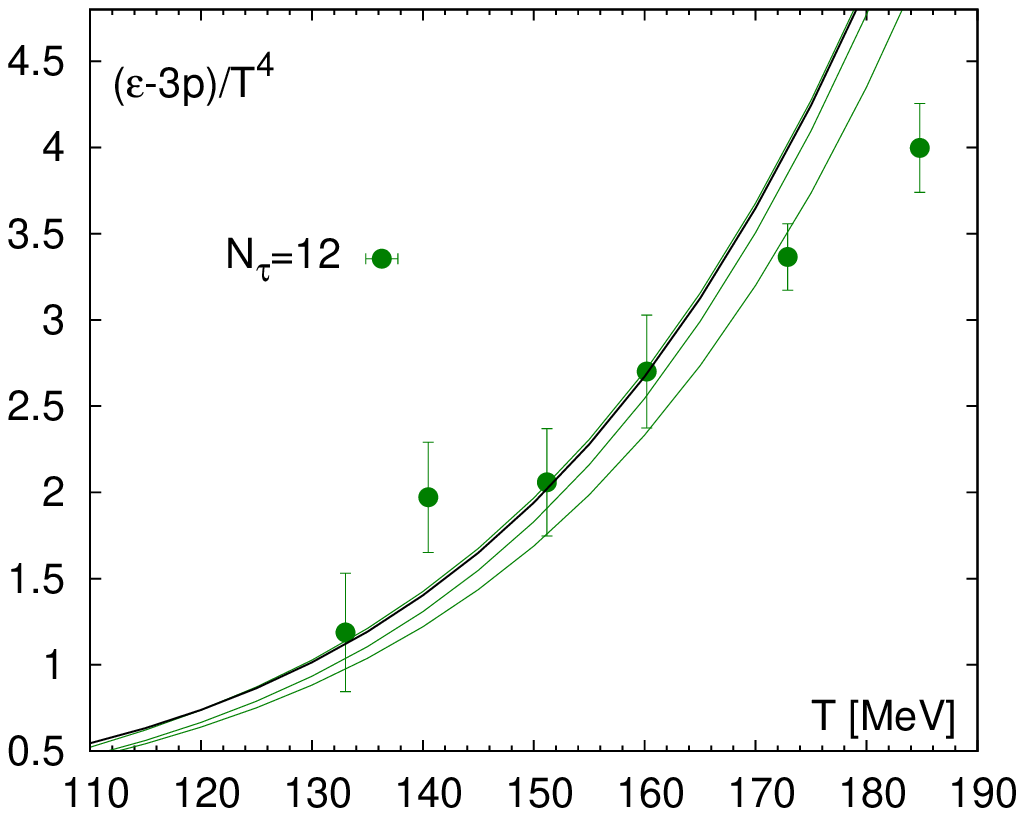}
\caption{The trace anomaly calculated in the HRG-QM and in the HRG-QM with 
modified hadron spectrum for $N_{\tau}=6,~8,~10$ and $12$, and compared with 
the lattice results, see Section \ref{secTmumu} for details.
The solid thick line corresponds to the continuum HRG-QM, while the top, 
middle and bottom colored thin lines correspond to the lattice HRG-QM, where 
only the ps mesons are modified, all ground states hadron are modified, and
all ground state and excited state hadrons are modified, respectively.}
\label{fig:e-3p_hrgNt}
\end{figure*}

\begin{figure*}
\includegraphics[width=7cm]{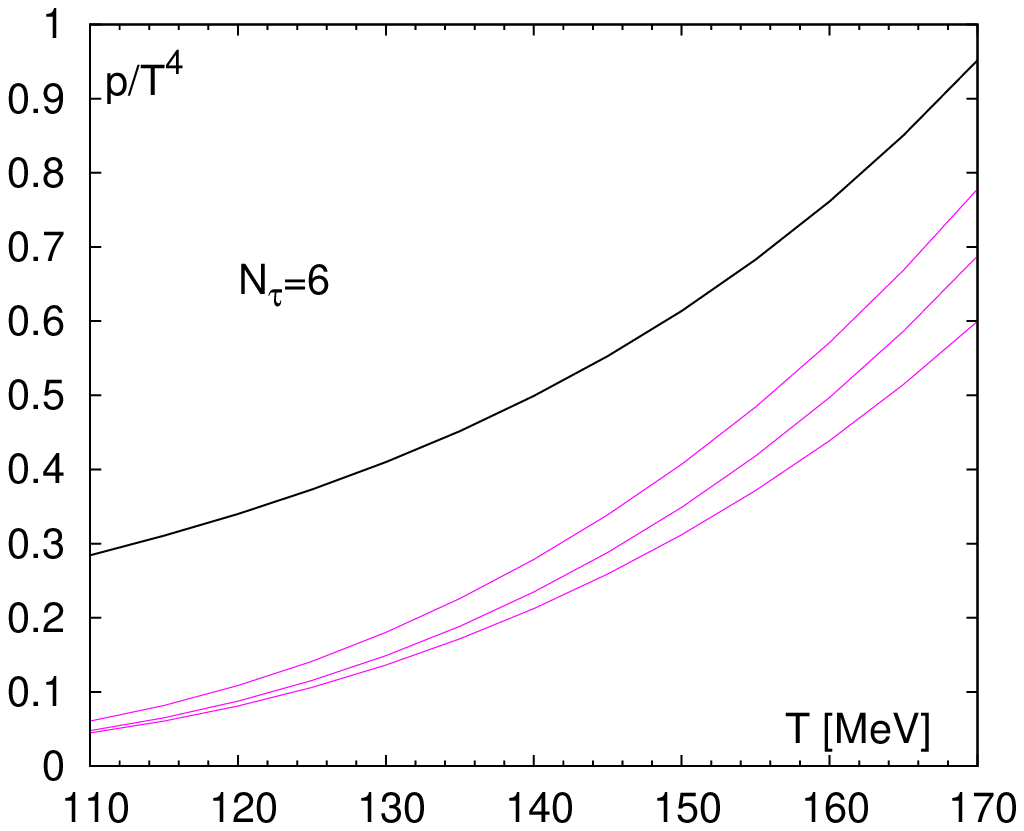}
\includegraphics[width=7cm]{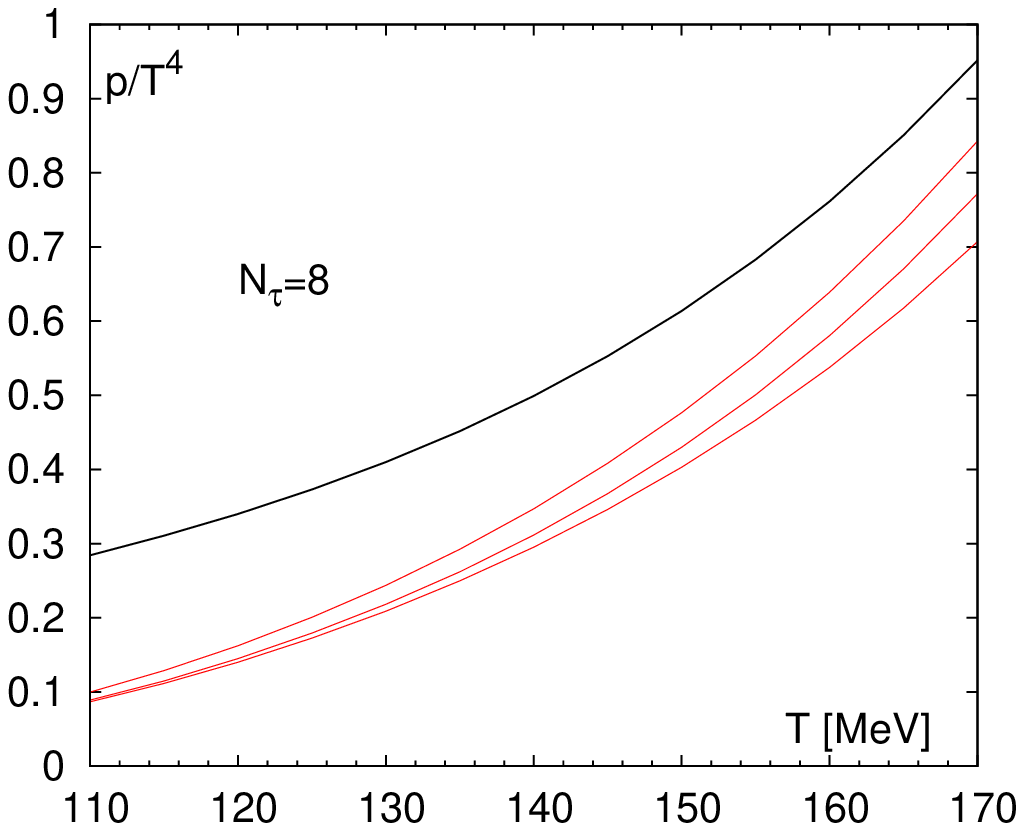}
\includegraphics[width=7cm]{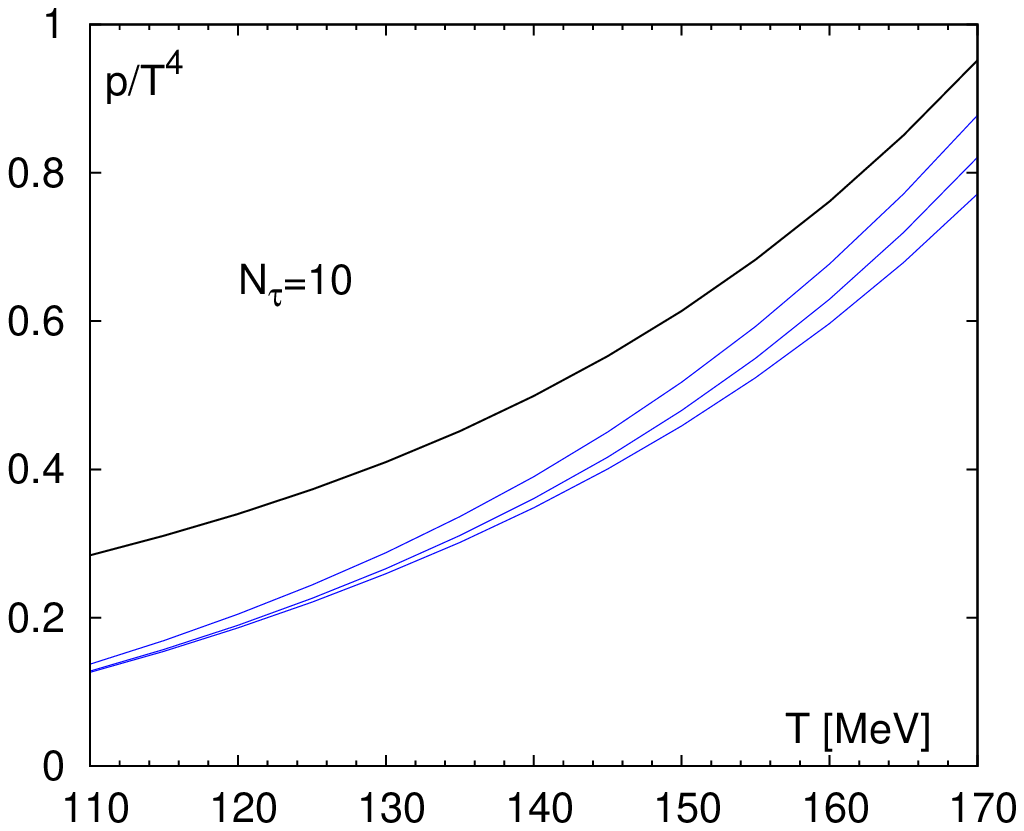}
\includegraphics[width=7cm]{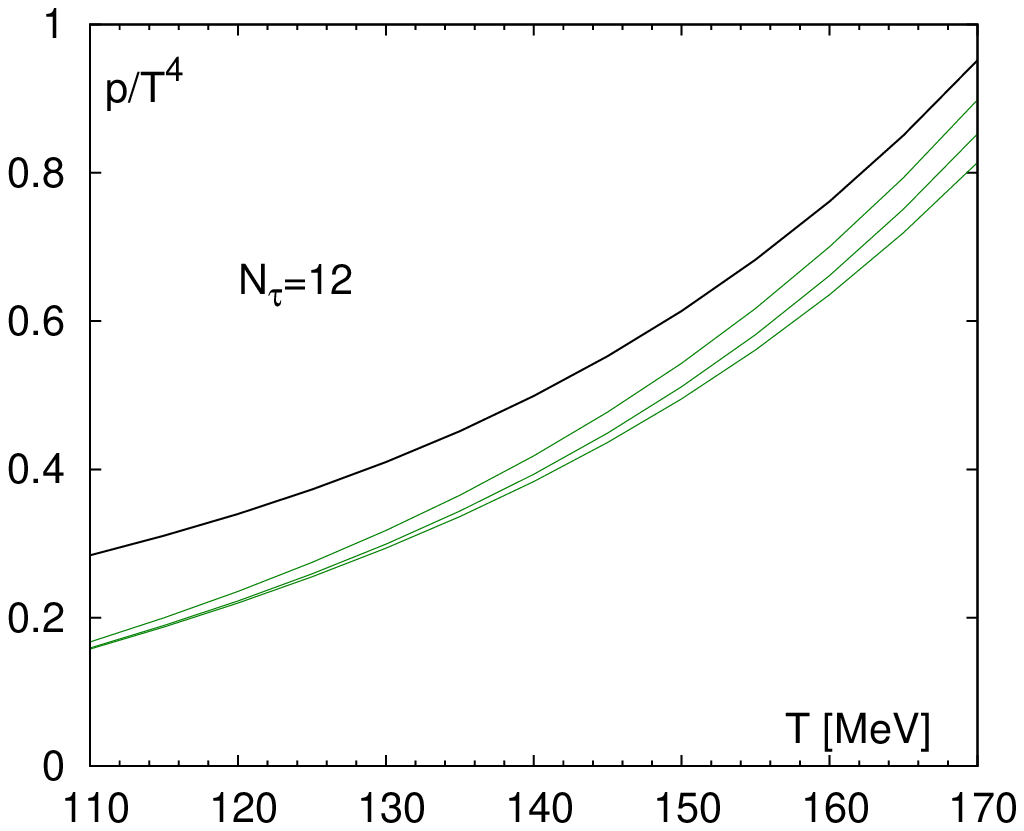}
\caption{The pressure calculated in the HRG-QM and in the HRG-QM with 
modified hadron spectrum for $N_{\tau}=6,~8,~10$ and $12$ from top left 
to bottom right.
The solid thick line corresponds to the continuum HRG-QM. The top, 
middle and bottom colored thin lines correspond to the lattice HRG-QM, where only 
the ps mesons are modified, all ground states hadron are modified and all the ground 
state and excited state hadrons are modified, respectively.}
\label{fig:p_hrgNt}
\end{figure*}

We need to consider also the contributions from the excited mesons and 
baryons. 
Unfortunately not much is known about the cutoff dependence of the excited 
hadron states in the staggered fermion formulations. 
We will work with two extreme assumptions about the cutoff dependence of 
the excited hadron states. 
First, we will assume that the masses of excited hadron states are not 
affected by the lattice cutoff. 
Second, we will assume that the masses of the excited hadron states are 
affected by the lattice cutoff the same way as the masses of the 
corresponding ground state hadrons.
Furthermore, we will calculate the EoS in the HRG model assuming that only 
ps mesons are affected by the taste-symmetry breaking. 
We will compare these three scenarios with the continuum HRG model
in order to understand the size of the cutoff effects. 
We will use the HRG model with missing states (HRG-QM) in what follows.
The trace anomaly calculated for different $N_{\tau}$ is shown in 
Fig. \ref{fig:e-3p_hrgNt} and compared to the lattice results. 
In Fig. \ref{fig:p_hrgNt} we show the pressure calculated for the same set 
of $N_{\tau}$ values.
We see that the difference between the continuum HRG and the lattice HRG is 
larger for the pressure than for the trace anomaly, and the continuum limit 
is approached from below. 
The difference in the cutoff dependence of the trace anomaly and the pressure
can be understood as follows. 
The cutoff effects make the hadrons heavier. 
This reduces the pressure as expected. 
However, states with larger masses contribute more to the trace anomaly. 
So this partially compensates the exponential suppression due to larger
quark masses in the case of the trace anomaly in the considered temperature 
range. 
At sufficiently low temperatures, the cutoff dependence of the pressure and 
the trace anomaly are qualitatively similar. 
We also note that the reduction of the pressure relative to the continuum 
HRG expectation is mostly due to the ps meson sector. 
As one can see from Fig. \ref{fig:p_hrgNt} taking into account
the modification of the baryon and vector meson masses in the HRG 
calculations only results in relatively small effects.
We also note that for p4 and asqtad actions the cutoff effects due to 
taste-symmetry breaking are much larger \cite{Huovinen:2009yb}.

We use the value of the pressure in the modified HRG-QM, in which the cutoff 
dependence of all the ground state hadrons is taken into account as discussed 
above (middle curves in Fig. \ref{fig:p_hrgNt}) to determine the pressure at 
some initial value of the temperature $T_0$ in the integral method (see 
Section \ref{secp}).
To estimate the uncertainty in $p(T_0)$ we consider the difference between 
the HRG-QM model in which only ps mesons are modified (upper curves in 
Fig. \ref{fig:p_hrgNt}) and the HRG-QM model in which all ground state 
and excited state hadrons are modified (lower curves in 
Fig. \ref{fig:p_hrgNt}).
The resulting values are given in Table \ref{tab:p0}.
\bibliography{ref}

\end{document}